\documentclass[smallextended,envcountsect]{svjour3}

\usepackage[english]{babel}
\usepackage{amsmath}
\usepackage{amssymb}
\usepackage{amsfonts}
\usepackage{mathrsfs}
\usepackage[utf8]{inputenc}
\usepackage{tabularx}
\usepackage{xcolor}
\usepackage{tikz}
\usepackage[hidelinks]{hyperref}
\usepackage[nameinlink, capitalise, noabbrev]{cleveref}

\newcommand{\comment}[1]{}

\newcommand{\mult}[1]{\mathtt{#1}}
\newcommand{\sk}{\mult{sk}}
\newcommand{\pk}{\mult{pk}}
\newcommand{\hp}[1]{\mult{HP}^{(#1)}}
\newcommand{\Mp}[1]{\mult{MP}^{(#1)}}
\newcommand{\tH}{\mult{t_H}}
\newcommand{\Z}{\mathbb{Z}}
\newcommand{\p}[1]{\mult{P}^{(#1)}}
\newcommand{\hatp}[1]{\hat{\mult{P}}^{(#1)}}
\newcommand{\pp}{\mult{PK}}
\newcommand{\vv}[2]{\mathbf{v}_{#1}^{(#2)}}
\newcommand{\oo}{\mathbf{O}_i}
\newcommand{\ff}{\mathbf{f}_i}
\newcommand{\gc}{\mathbf{g}_i}
\newcommand{\TT}[2]{\mathbf{\Theta}_{#1}^{(#2)}}
\newcommand{\TTT}{\mathbf{\Theta}}
\newcommand{\mm}[2]{\mult{m}_{#1}^{(#2)}}
\newcommand{\sig}[2]{\mult{sig}_{#1}^{(#2)}}
\newcommand{\ones}{\mathbf{1}}
\newcommand{\zeros}{\mathbf{0}}

\newcommand{\expected}{\mathbb{E}}
\newcommand{\probability}{\mathbb{P}}

\newcommand{\step}[2]{
    \vspace*{5pt}
    \noindent\rule{\textwidth}{1pt}
    \vspace*{-15pt}
    \paragraph{STEP #1}\emph{(#2)}
}
\newcommand{\checkstep}[1]{
    \vspace*{0pt}
    \noindent\rule{\textwidth}{1pt}
    \vspace*{-15pt}
    \paragraph{#1}~\\
    
    \noindent
}

\newcounter{protocol}

\spnewtheorem{observation}{Observation}[section]{\bfseries}{\itshape}

\spnewtheorem{thm}{Theorem}[section]{\bfseries}{\itshape}
\crefname{thm}{theorem}{theorems}
\Crefname{thm}{Theorem}{Theorems} 

\spnewtheorem{cor}[section]{Corollary}{\bfseries}{\itshape}
\crefname{cor}{corollary}{corollaries}
\Crefname{cor}{Corollary}{Corollaries}

\spnewtheorem{prop}{Proposition}[section]{\bfseries}{\itshape}
\crefname{prop}{proposition}{propositions}
\Crefname{prop}{Proposition}{Propositions}

\spnewtheorem{lem}{Lemma}[section]{\bfseries}{\itshape}
\crefname{lem}{lemma}{lemmas}
\Crefname{lem}{Lemma}{Lemmas}

\crefname{enumerate}{item}{items}
\Crefname{enumerate}{Item}{Items}

\spnewtheorem{prob}{Problem}[section]{\bfseries}{\itshape}
\crefname{prob}{problem}{problems}
\Crefname{prob}{Problem}{Problems}

\definecolor{lime}{HTML}{A6CE39}
\DeclareRobustCommand{\orcidicon}{%
	\begin{tikzpicture}
	\draw[lime, fill=lime] (0,0) 
	circle [radius=0.16] 
	node[white] {{\fontfamily{qag}\selectfont \tiny ID}};
	\draw[white, fill=white] (-0.0625,0.095) 
	circle [radius=0.007];
	\end{tikzpicture}
	\hspace{-2mm}
}

\foreach \x in {A, ..., Z}{%
	\expandafter\xdef\csname orcid\x\endcsname{\noexpand\href{https://orcid.org/\csname orcidauthor\x\endcsname}{\noexpand\orcidicon}}
}

\makeatletter
\newcommand{\labelx}[1]{
    \relax
    \ifmmode
        \label{#1} 
    \else 
        \ifnum\pdfstrcmp{\@currenvir}{document}=0
            \label{#1}
        \else
            \label[\@currenvir]{#1}
        \fi
    \fi
}
\makeatother

\title{Cob: a Leaderless Protocol for Parallel Byzantine Agreement in Incomplete Networks}
\titlerunning{Cob: a Leaderless Protocol for Parallel Byzantine Agreement}

\author{Andrea Flamini \and Riccardo Longo \and Alessio Meneghetti}

\authorrunning{A. Flamini, R. Longo, A. Meneghetti}
\institute{
    Department of Mathematics, University Of Trento, 38123 Povo, Trento, Italy\\
    A. Flamini\hspace*{\fill}
    \email{andrea.flamini.1995@gmail.com}\\
    \orcidA{} https://orcid.org/0000-0002-3872-7251\\
    R. Longo\hspace*{\fill}
    \email{riccardolongomath@gmail.com}\\
    \orcidB{} https://orcid.org/0000-0002-8739-3091\\
    A. Meneghetti\hspace*{\fill}
    \email{alessio.meneghetti@unitn.it}\\
    \orcidC{} https://orcid.org/0000-0002-5159-7252
}
\begin{document}

\maketitle

\begin{abstract}
    In this paper we extend the \emph{Multidimensional Byzantine Agreement (MBA) Protocol},
    a {leaderless} Byzantine agreement for lists of arbitrary values, into a protocol suitable for wide gossiping networks: \emph{Cob}.
    
    This generalization allows the consensus process to be run by an incomplete network of nodes provided with (non-synchronized) same-speed clocks.
    Not all nodes are active in every step, so the network size does not hamper the efficiency, as long as the gossiping broadcast delivers the messages to every node in reasonable time.
    These network assumptions model more closely real-life communication channels, so Cob may be applicable to a variety of practical problems, such as blockchain platforms implementing sharding.
    
    Cob has the same Bernoulli-like distribution that upper-bounds the number of steps as the MBA protocol. We prove its correctness and security assuming a supermajority of honest nodes in the network, and compare its performance with Algorand.
\end{abstract}


\paragraph{Acknowledgments}
The core of this work is contained in the first author's MSC thesis.
Part of the results presented here have been carried on within the EU-ESF activities, call PON Ricerca e Innovazione 2014-2020, project Distributed Ledgers for Secure Open Communities.
The authors are members of the INdAM Research group GNSAGA.
We would like to thank the Quadrans Foundation for their support.
The authors would like to thank the anonymous referees.

\section*{Declarations}
The authors have no relevant financial or non-financial interests to disclose.
The authors have no conflicts of interest to declare that are relevant to the content of this article.
The authors have no financial or proprietary interests in any material discussed in this article.

\section{Introduction}
\labelx{intro}
One of the main issues that blockchain platforms must deal with is the lack of \emph{scalability}.
Scalability is the ability of a platform to grow and manage an increasing number of requests.
In particular, we say that a blockchain \emph{scales} if it can easily adapt to changes in the number of users that decide to join in, as well as in the number of transaction requests that such users broadcast to the nodes maintaining the blockchain platform. In order to solve the scalability issues of blockchain platforms, many approaches have been proposed over the years. Some of them are the \emph{block size increase}, the use of \emph{off-chain state channels}, \emph{segregated witness (SegWit)} and \emph{sharding}.
Among the proposed approaches, sharding seems to be the most promising \cite{meneghetti2019survey}.

The term sharding comes from database management, where it identifies a particular type of database partitioning, that consist in dividing large databases into smaller parts, called shards.
Shards are more manageable in terms of server hosting and other aspects of database maintenance, and allow to have faster query time by diversifying the responsibility of a database structure.

In the context of blockchain design, sharding consists in breaking the blockchain into small parts that are managed in parallel by node subsets, called \emph{shards}. This augments throughput, since many transactions can be simultaneously validated, allowing blockchains to effectively scale for a huge number of users. Many blockchain platforms use sharding as a mean to reach scalability, for example Ethereum 2.0 \cite{buterin2014next}, Zilliqa \cite{team2017zilliqa} or EOS \cite{io2017eos}.
For a better description of blockchain sharding we refer to the survey of Meneghetti et al. \cite{meneghetti2019survey}. Together with the security concerns regarding how to distribute the nodes among different shards (so that groups of cooperating malicious nodes are not assigned to the same shard), one of the main issues the protocol designer must deal with is the way the transactions validated by different shards can reach compatibility with one another.
This problem is also referred to as \emph{reconciliation problem}.

Since blockchains are distributed ledgers, one of the core component is represented by the consensus protocol that the blockchain network must execute.
The consensus protocol allows the nodes of the network to update their copy of the ledger in the very same way.
Therefore, a blockchain implementing sharding must be able to bring the network to agreement (reach consensus) on which blocks are legitimately created by each shard, so that the network can proceed with the reconciliation of such transactions.
One possible way to achieve this goal is to let the network execute a consensus protocol and decide which shards legitimately produced a valid block.

\subsection{Cob Protocol}
In this paper we present \emph{Cob}, a novel consensus protocol which efficiently solves the problem of reaching consensus on a set of blocks legitimately created by each shard.
In particular we propose a viable solution for permissionless blockchain networks. 

This problem can be easily extended to the following more general problem:
\begin{prob}\labelx{general_problem}
    Given a set of events which a network of nodes can observe, how can the nodes reach consensus on some relevant information about such events?
\end{prob}

In the context of blockchains implementing sharding the events to be observed are the creation and diffusion of a block by each shard; the relevant information about the event is the content of the block or some data which identifies it (e.g. the digest of the block computed via an hash function). In this paper we will describe Cob following the more general problem (i.e.  \Cref{general_problem}), however, the reader can keep in mind the specific application of reaching consensus over the shards. 

Given a set of $m$ events the nodes can observe, an instance of Cob requires every node in the network to build a list with $m$ components.
Each event (e.g. creation of a block of a shard) is associated to a component of the list and, once the nodes observe an event, they locally record in the corresponding component the relevant information about such event (e.g. the digest of the newly created block).
This can be referred to as the \emph{observation phase}.

After the observation phase, the nodes of the network will continue with a Cob protocol execution, exchanging messages until they reach consensus on a list of relevant information. 

\subsubsection{Cob: a Parallel Consensus Protocol}
\labelx{parallel}
We say that Cob is a \emph{parallel} protocol since it is designed in a way that the consensus process is carried out simultaneously on each component of the list by every node involved in the consensus protocol.
Every message broadcast by the nodes contains some information about each component, but the consensus achievement on each component is independent from the others.

In particular the nodes will exchange lists of values during the whole protocol execution.
In the first 3 steps they will exchange lists of strings (the relevant information) and in the following steps they will exchange lists of bits in order to reduce the bandwidth required.

Agreement might be reached faster in some components, however the nodes will stop the protocol execution only when they realize that they agree on every component, therefore on the whole list.

If an event is not detected (e.g. a shard did not broadcast any block) or it is impossible to reach consensus (e.g. multiple blocks have been broadcast by the same shard), agreement will be reached on the special value $\bot$.\\

It is crucial that agreement is carried out in parallel on the list components instead of on the whole list, otherwise a widespread disagreement on a single component would affect the consensus achievement on the other components as well.
Conversely, if agreement is carried out in parallel and independently, it is possible to preserve and finalize the agreed upon components, and to set to $\bot$ the controversial components.

\subsubsection{Cob: a Leaderless Consensus Protocol}
\labelx{leaderless}
Since in permissionless blockchain networks anyone can join the network, we must consider an attacker that may try to disrupt the consensus process in several ways.
For example, if we stick to the example of the blockchain implementing sharding, when an attacker shares with the network the list built during the observation phase, it may broadcast different lists to different nodes, advertising different blocks related to the same shard.
This would cause a network partition in groups with different views about what the attacker has seen.
Another attack it may perform is a censorship attack, pretending not to have received a block by a specific shard, setting the related component to $\bot$.

Many classical consensus protocols used by blockchain networks to record transactions \cite{nakamoto2008bitcoin,wood2014ethereum,chen2019algorand,team2017zilliqa,kiayias2017ouroboros} (e.g. cryptocurrency transfers, smart contract execution requests) do not guarantee that a specific transaction will be included into the blockchain as soon as it is broadcast, in the newly created block.
In many cases this is just fine, in fact the transaction can be included into one of the following blocks after waiting a reasonable time interval. 

For this reason, many consensus protocols are \textit{leader based}, which means that there is a node which proposes a new block, and then the network decides whether to accept it or not.
If the network decides to discard the leader proposal it elects a new leader and starts over. 
However, it is essential that the leader must change from time to time so that, if an attacker tries to undermine the liveness of the platform or to practice censorship, a (eventually honest) new leader will propose a new block (in one case), or will include the deliberately excluded transactions (in the other case).

The \Cref{general_problem} differs from the problem of (eventually) recording transaction requests.
The events observed by the network must be discussed right after they are observed and the relevant information must be included in the agreed upon list or excluded once and for all.

Therefore, it is essential for our consensus protocol to be \emph{leaderless}.
In fact, a leader, if honest, would propose a list of relevant information which is heavily influenced by their own point of view (which in some cases might lead them to take incorrect decisions), and if the leader is malicious may easily perform censorship attacks refusing to include some information in the list, or deliberately include invalid information.
In both cases, if the network does not agree even with one component proposed by the leader, it will reject the leader proposal, and this process is repeated until a leader proposes a list which gets accepted by a majority of the network.
Note that this might not even happen, in fact, if there is a wide disagreement among the nodes about one or more components, there might not exist a list which is accepted by such majority. 

So we have shown that there are many reasons that suggest to abandon the leader-based approach in favour of a leaderless approach.
 A leaderless consensus protocol, instead of questioning a single node, requires several nodes to share their own proposal.
 Then, based on these proposals, the protocol will bring the network to a consensus on a shared output as we will show in \cref{protocol description} and prove in \Cref{security}.

We will show in \Cref{protocol description} that in the first step of Cob, a set of randomly chosen nodes will share with the network their observed values, namely the list of relevant information (e.g. the hash of the shard blocks they have received).
Starting from this information, the network will carry on with the consensus protocol to decide which components of the list can have a value, since the nodes of the network agree on some relevant information, and which will be set to $\bot$.

\subsection{State of the Art and Considerations}
\labelx{stateofart}
In \cite{flamini2021multidimensional} is presented the MBA Protocol, a solution to \Cref{general_problem} for a relatively small network of a fixed number $n$ of nodes, under some strong communication assumptions, and under the threat of an attacker that controls less than $\frac{1}{3}$ of the nodes.
In particular is assumed a strongly synchronous communication model and a complete network, where every node could instantaneously send a message to each other.
These assumption are unrealistic or dramatically reduce the possible application contexts.

In this paper we go a step further and define an analogous protocol called Cob, which works under more realistic assumptions and can be executed by a network of nodes of any size and the nodes communicate gossiping the messages broadcast in the network.
This model makes Cob a viable solution to \Cref{general_problem} for a permissionless blockchain platform network.
In fact, as we will see in \Cref{netass}, every step of the protocol is executed by a randomly selected set of nodes, whose cardinality is constant in expected value: this guarantees that, independently of the size of the network, the number of messages broadcast during each step will be constant.\\

As we have stated in \Cref{leaderless}, we are interested in leaderless consensus protocols which work under realistic network and communication assumptions. 

There is no doubt that \emph{asynchronous} BFT protocols would be the best solution for building high-assurance and resistant consensus protocols.
Unlike \emph{synchronous} or \emph{weakly synchronous} protocols, whose liveness relies on communication assumptions, the asynchronous protocols do not put their liveness at risk.

In 2016, Miller et al. presented the leaderless asynchronous BFT protocol HoneyBadgerBFT \cite{miller2016honey}, which significantly improved prior asynchronous BFT protocols \cite{ben1994asynchronous,cachin2001secure,cachin2002secure}. 
HoneyBadgerBFT is based on Asynchronous Common Subset (ACS) \cite{ben1994asynchronous} implemented in combination with the asynchronous binary consensus protocol of Mostéfaoui et al. \cite{mostefaoui2014signature}.
HoneyBadgerBFT can be used as consensus protocol for blockchains, achieving a throughput of 200 KB/s of data appended to the ledger using 10 MB blocks (therefore it requires 5 minutes for each protocol run) using 104 participating servers.
Later, in 2018, Duan et al. improved the HoneyBadgerBFT protocol presenting BEAT \cite{duan2018beat}, a family of five asynchronous consensus protocols designed to meet different goals, such as different performance metrics (scalability, bandwidth or latency).
Another improvement to HoneyBadgerBFT was proposed by Guo et al. with the Dumbo protocol \cite{guo2020dumbo}.

However, these leaderless asynchronous BFT protocols have performance (latency, throughput) issues when they are executed by a number of replicas which exceeds the hundreds.
For this reason, these protocols, or their variations, could be adopted only by permissioned or private blockchain platforms, because they are often controlled by relatively few nodes, but they do not provide a viable solution to the consensus problem in the context of permissionless networks.
Moreover, these protocols must be executed by a fixed set of nodes who actively partake to the communication protocol.
Therefore, these protocols can be incapable to guarantee resistance to targeted attacks that either compromise the servers involved or disconnect them from the network. \\ 

Since in this paper we are interested in solving a problem which can be applied to permissionless blockchain platforms implementing sharding, we assume that the number of nodes in the network can grow with no limit (potentially reaching millions of nodes) still guaranteeing the highest level of decentralization. 

Therefore the asynchronous protocols mentioned above, classical primary-backup protocols such as PBFT \cite{castro1999practical} and other recent concurrent protocols such as Mir-BFT \cite{stathakopoulou2019mir} or RCC \cite{gupta2021rcc} must be considered impractical for our use case due to the reduced and fixed number of actors involved in the protocol execution. 

For this reason we must make some compromises, and find a solution which allows the implementation of a scalable platform settling for a protocol which relies on assumptions that are as weak as possible.\\

In this regard, Algorand \cite{chen2019algorand} is a blockchain platform for cryptocurrency which adopts a BFT consensus protocol that faces three main challenges: 
\begin{enumerate}
    \item \textbf{avoid Sybil attacks}: this is done using \emph{weighted users}.
    Every user is weighted based on the amount of money in their account, therefore as long as more than $\frac{2}{3}$ of the money is in honest hands, then the protocol is proven to be secure. 
    \item \textbf{can scale to millions of users}: this is achieved by choosing for every step a different committee, a small set of representative randomly selected from the set of users based on the users' weight.
    \item \textbf{resists to denial of service attacks or disconnection of users performed by an attacker}: in fact, relying on a committee which performs the operations, gives the possibility of targeted attacks against the chosen committee members. 
    To prevent this, the protocol selects the committee members in a private and non interactive way via a verifiable random function (VRF) on the users' private key and some public data from the blockchain, a technique pioneered by Rabin in \cite{rabin1983randomized} which simulates a random lottery. 
    
    Once a player realizes they are selected in a committee then they must broadcast a message containing, among other things, the proof of their selection.
    At this point the attacker clearly knows about the selection and can try to corrupt such player.
    However, the nodes executing the protocol randomly change at every step and the attacker can not know in advance which node will be selected to broadcast a new message.
    This property is called \emph{player replaceability} and protects the network from targeted attacks. 
\end{enumerate}

All these properties are achieved assuming that the attacker can corrupt or take control of any user in the network but, as we mentioned before, more than $\frac{2}{3}$ of the money (on which each user's weight is based) must always be in honest hands.\\

Algorand, to guarantee liveness, makes a strong synchrony assumption \cite{gilad2017algorand} requiring that almost every honest node (e.g. 95$\%$), when it broadcasts a message, reaches almost every other honest node (e.g. 95$\%$) within a predetermined time interval.
The time is measured by each node by using \emph{same speed clocks} which might not be synchronized (pointing to different times) as long as they have the same speed.

Moreover, to guarantee safety a weak synchrony assumption is used: the network can be asynchronous (i.e. controlled by an adversary) for a long period of time, as long as this time is bounded (e.g. 1 day, 1 week).
After that the network must be synchronous for a reasonably long period of time (e.g. few hours, 1 day) in order to ensure safety \cite{gilad2017algorand}.

In this paper we build on top of Algorand's network and communication model a consensus protocol which solves the problem of reaching consensus on a list of relevant information about some observed events.
In fact, for our scope, it is essential to involve a high number of nodes to safely determine that this information has been correctly recorded, so we cannot rely on asynchronous protocols (due to their limitations) and must compromise on the network assumptions.

\paragraph{Outline}
In \Cref{preliminaries} we establish the preliminaries necessary to describe Cob.
We define our network assumptions, we recall some useful notation, then we describe the sortition mechanism that selects which nodes are active in each step of the protocol, giving the necessary definitions.
Finally we describe our assumptions on the honesty of the nodes.

In \Cref{protocol} we introduce the actual protocol, presenting a reference list of all the parameters and then describing in detail every step.

Then, in \Cref{security} we formally analyze the properties of Cob, proving that it is a Byzantine agreement through a series of preparatory lemmas and propositions.
The main theorem also gives a probabilistic upper bound on the number of steps that are necessary to halt the execution.

In \Cref{performance} we analyze the message complexity and the weight of the data broadcast in the network in each protocol execution.
We also compare the performance of Cob with the one of Algorand's consensus protocol \cite{chen2019algorand} in solving \Cref{general_problem} in the context of blockchain platforms implementing sharding.

Finally in \Cref{conclusions} we draw some conclusions and remarks, and outline future works to improve the applicability of the protocol.

\section{Preliminaries}
\labelx{preliminaries}
In this section we define some assumptions, preliminary concepts, and notations that will be used later on to describe Cob and prove its properties.

\subsection{Network Assumptions}
\labelx{netass}
In complete networks, the number of messages exchanged through the network grows exponentially with the number of network participants, so for practical applications it is more convenient to consider a different network model, such as the \emph{Asynchronous Gossiping Network} (AG networks)\footnote{ Algorand describes the environment in which is defined as asynchronous. This is because the communications between nodes happen via gossip and the protocol steps, which for a single user are non-overlapping time intervals, for different users may overlap due to asynchrony.
However, since Algorand assumes that exists a predetermined upper-bound to the time required by a message to reach (almost) every node, and therefore exists an upper-bound to the delay between different nodes, Algorand can not be considered an asynchronous protocol. } presented by Micali in Algorand \cite{chen2019algorand}.

In this model messages are broadcast in the network in a gossiping fashion: a procedure characteristic of peer-to-peer communications where messages pass from one node to its neighbours and so on until they reach every node.
In gossiping networks we rely on each member to pass messages along to its neighbours, therefore it is reasonable to envisage the network as an incomplete, connected and non-directed graph.
We assume that a message sent by an honest node reaches every honest node within a time limit that depends on the size of the message itself.
Since malicious nodes can behave arbitrarily, this assumption means that malicious nodes cannot be cut vertices in the network graph, that is the graph remains connected even without the edges connected to malicious nodes.
We will also require that the ratio of malicious or faulty nodes is less than $\frac{1}{3}$.

In an AG network there does not exist a common clock, but we assume that all  network participants are provided with \emph{Same-Speed Clocks} \cite{chen2019algorand}.
In other words, we assume that each network participant has its own clock and that the clocks all have the same speed, even if they are not synchronized in any other way.

We take as time frame reference the earliest clock in the network, and suppose that the protocol execution starts at time 0, i.e. time starts when the first player begins the protocol execution.
Moreover we assume that the discrepancy between any two clocks is at most a constant $\lambda$, that also upper bounds the time required to diffuse a ``short'' message of the protocol to the whole network (see \Cref{paralist}), so each player will start the execution of the protocol at a time comprised in the interval $[0, \lambda]$.
For example, this discrepancy could be observed in a scenario in which the first player triggers the start of the protocol execution, broadcasting a signal and resetting its clock, and then each player starts the execution (and resets its clock) when it receives this signal, with the network delay causing the discrepancies.
Afterwards the time discrepancies do not vary because of the same-speed nature of the clocks.

\subsection{A Cryptographic Sortition Mechanism}
In the protocol described in \Cref{protocol description} not every player in the network is always active (i.e. authorized to broadcast messages), on the contrary at every step some players are selected to be active, while the others have a passive role.
In order to better clear up this distinction, from now on in a specific step we will call \emph{players} only the nodes selected to be active and broadcast their message, while a generic node of the network will be referred as a \emph{user}.
We will denote with $\p s$ the set of players of step $s$.

We want this selection to be random, and furthermore we would like it to be \emph{private} and performed without the aid of a trusted third party.
With private we mean that each user should be able to privately check if it will be selected to be active (i.e. a player) in a step, and then be able to prove its selection to the other players.
This concept is closely related to that of \textit{verifiable random functions} (VRF), i.e. pseudo-random functions which provide publicly verifiable proofs of their outputs' correctness.

In our protocol the sortition is implemented through a cryptographic hash function $H$ (modeled as a random oracle) and a digital signature scheme $\left(G, S, V\right)$ with the uniqueness property, which is defined as follows.
\pagebreak
\begin{definition}[Digital Signature Scheme with Unique Signature]
\labelx{unisig}
    A digital signature scheme with unique signature is a triple of algorithms $\left(G, S, V\right)$ such that:
    \begin{itemize}
        \item $G$ is the key generation algorithm that outputs a secret key $\sk$ and a public key $\pk$;
        \item $S$ is the signing algorithm, that given a message $m$ and a private key $\sk$ outputs a signature $\sigma = S\left(\sk, m\right)$;
        \item $V$ is the verification algorithm that given a message $m$, a signature $\sigma$ and a public key $\pk$ outputs either $\mult{true}$ or $\mult{false}$, and such that:
        \begin{itemize}
            \item the scheme is correct, i.e. for every $\left(\sk, \pk\right)$ generated with $G$ it holds
            \begin{equation*}
                V\left(\pk, S\left(\sk, m\right), m\right) = \mult{true} \quad \forall m\;;
            \end{equation*}
            \item the signature is unique, i.e. for any probabilistic polynomial time algorithm $F$ that given a message outputs a public key $\hat\pk$ and two distinct signatures $\hat\sigma \ne \tilde\sigma$, we have that:
            \begin{equation*}
                \mathbb{P}\left(V\left(\hat\pk, \hat\sigma, m\right) = V\left(\hat\pk, \tilde\sigma, m\right) = \mult{true}\right) < \varepsilon \quad \forall m
            \end{equation*}
            where $\varepsilon$ is negligible.
            Note that this property holds also for public keys whose relative private key is known, and even for values $\hat\pk$ that are not legitimately generated public keys.
        \end{itemize}
    \end{itemize}
\end{definition}

Given this definition, we can now describe the sortition of the active players in each step of the protocol.

\begin{definition}[Sortition Mechanism]
    Let $\left(G, S, V\right)$ be a digital signature scheme with unique signature, and suppose that every user $1\le i \le N$ is identified by a public key $\pk_i$, let $r$ be a random string independent from $\pk_i$ for every $1\le i \le N$, and suppose that every user knows $r$ and $\{\pk_i\}_{1\le i \le N}$.
    Moreover let $n\le N$ be the desired number of players during each step $s$ of the protocol, let $H:\{0,1\}^* \longrightarrow \{0,1\}^d$ be a hash function, and let ${\phi: \{0,1\}^d \longrightarrow (0,1]}$ be the standard decoding of a bit string into the unit interval $\phi\left(h\right) = \frac{1+\sum_{i=0}^{d-1} h_i 2^i}{2^d}$.
    
    User $i$ is selected to be a player during step $s$ of the protocol, i.e. $i \in \p s$, if:
    \begin{equation*}
        \phi\left(H\left(\sigma_i^{\left(s\right)}\right)\right) \le \frac{n}{N} \quad \bigwedge \quad V\left(\pk_i, \sigma_i^{\left(s\right)}, H\left(r\|s\right)\right) = \mult{true}\;.
    \end{equation*}
    Where $\sigma_i^{\left(s\right)}= S\left(\sk_i, s\|r\right)$.
    The signature $\sigma_i^{\left(s\right)}$ can then be used to prove that $i \in \p s$.
\end{definition}

Note that, when $H$ is modeled as a random oracle, $\phi\big(H\big(\sigma_i^{(s)}\big)\big)$ is uniformly distributed, so the probability of a player to be selected is $\frac{n}{N}$, and the expected number of active players is indeed $n$.

Note that the same sortition mechanism can be implemented with a weaker notion of signature scheme that allows signing failures.
That is, a scheme where the output of $S$ is a special symbol $\bot$ with fixed probability $f$ (supposing the message to be uniform in $\{0,1\}^d$), with ${V\left(\pk, \bot, m\right) = \mult{false}}$ for every $\pk$ and $m$.
The sole adjustment required is to increase the threshold $\frac{n}{N}$ to account for the signing failure, using $\frac{n}{N \left(1-f\right)}$ instead.

In some applications it might be desirable that players are selected with nonuniform probability.
There is a simple trick to adjust the selection probability for each player: let $p\in(0,1]$ be a fixed probability and $t_i\ge 1$ be a publicly known threshold for player $i$.
The tweaked process selects player $i$ if it can provide a pair signature-counter $\left(\sigma_i^{\left(s\right)}, c_i\right)$ such that $\phi\left(H\left(\sigma_i^{\left(s\right)}\right)\right) \le p$, $V\left(\pk_i, \sigma_i^{\left(s\right)}, H\left(r\|s\|c_i\right)\right) = \mult{true}$, and $c_i \le t_i$.
In other words the player $i$ has $t_i$ attempts to produce a \emph{winning} signature, so its probability to be selected is $1 - \left(1 - p\right)^{t_i}$.


\subsection{Sortition Assumptions}
Similarly to the MBA protocol \cite{flamini2021multidimensional}, Cob requires that the number of honest nodes at each step $s$ is more than two times the number of malicious players active at step $s$. 
Since the active players are randomly selected, we require that at each step there are enough active honest players with high probability.

We now define a probabilistic concept which will be widely used throughout this paper.
\begin{definition}
We will write that an event $E$ happens with \emph{overwhelming probability} if $P\left(E\right)\ge 1-\epsilon$, where $\epsilon\in \left(0,1\right)$ is a parameter sufficiently close to 0.
\end{definition}
A good choice for practical applications could be $\epsilon=10^{-12}$.\\

\begin{definition}
    \labelx{th_assumption}
    Let $n$ be the expected number of active players in a step, we define the threshold $\tH = \lfloor\frac{2n}{3}\rfloor+1$.
    For every step $s$ we choose the parameter $n$ in a way that the following relationships between the number of honest players $\hp s $ and the number of malicious players $\Mp s $ hold with overwhelming probability:
    \begin{enumerate}
        \item $| \hp s  |> \tH$;
        \item $| \hp s  |+2| \Mp s  |<2\tH$.
    \end{enumerate}
\end{definition}

Note that these two conditions imply that $\hp s >2\Mp s $.
In practice, they imply that with overwhelming probability:
\begin{itemize}
    \item the protocol has, at each step, the required $\frac{2}{3}$ honest majority of players.
    
    \item at every step there is a sufficient number of honest players who can certify a new list or finalize a list component;
    
    \item two distinct nodes can not finalize the same component with two distinct values.
\end{itemize}
Note that the closer to 1 the ratio of honest users in the network is, the smaller the number of players for each step needs to be. 

The parameter choice necessary to meet the requirements is done using variants of Chernoff bounds, as in Algorand \cite{chen2019algorand} and the analysis of such bound can be found in \cite{cai2019analysis}.

\subsection{Notation}
We will typeset lists in boldface and in general subscript will be used to denote the player who created the value and the index of list components, while superscripts will refer to the protocol step in which the value has been produced.
So $\vv i s$ will be a list created by user $i$ during step $s$ of the protocol, while $v_{i, c}^{\left(s\right)}$ will denote the $c$-th component of said list.
As shorthand, $\ones$ denotes a list where each component is equal to $1$, and similarly $\zeros$ denotes an all-zero list.

As in \cite{chen2019algorand} and \cite{flamini2021multidimensional}, the notation $\#_i^{\left(s\right)}\left(v,c\right)$, for $1 \le c \le m$ represents the number of players from which player $i$ has received during step $s$ a valid message containing a list $\vv{}{s}=\left(v_1,\dots,v_m\right)$ such that $v_c=v$ considering, possibly, also its own message.
We recall that honest players consider at most one message from player $j$ as valid (discarding all contrasting and not properly formatted messages, and counting identical messages as one), so only \emph{valid} messages are considered and counted, and $\sum_{v} \#_i^{\left(s\right)}\left(v, c\right)\leq | \p s | \quad \forall i, s, c$.

In the protocol the players try to reach agreement on a list of arbitrary values, where each player $j$ starts the protocol knowing an $m$-dimensional list $\mathbf{v}_j=\left(v_{j,1},\ldots,v_{j,m}\right)\in V = \prod_{c = 1}^m V_c$.
We say that the players have reached \emph{$c$-agreement}, where $1 \le c \le m$ is a specific component, when there exists $v \in V_c$ such that for every honest player $j$, $v_{j,c}=v$.
When $c$-agreement is reached on all the components of the list, we have that for all honest players $i,j$, $\mathbf{v}_i=\mathbf{v}_j$, hence also agreement is reached.

\section{Cob Protocol}
\labelx{protocol}
We now present Cob, a protocol that allows a wide gossiping network to reach agreement on a list of arbitrary values.
The properties of the protocol will be formally stated and proved in~\Cref{security}.

\subsection{Protocol Parameters and Components}
\labelx{paralist}
For the sake of clarity and easy reference, we now provide a list with the definition of the parameters and the notation that we will use to describe and analyze Cob:
\begin{itemize}
    \item $H$: a cryptographic hash function, modelled as a random oracle;
    
    \item $\left(G, S, V\right)$: a digital signature scheme with unique signature (see \Cref{unisig});
    
    \item $N\in \Z^+$: the number of nodes in the network, i.e. the users of the protocol;
    
    \item $\pp$: the set of public keys of the users, each user $i$, with $1 \le i \le N$, is univocally identified by its public key $\pk_i\in\pp$ and has a private key $\sk_i$;
    
    \item $ \frac{2}{3}< h \le 1$: the ratio of the honest users in $\pp$;
    
    \item $r$: a reference string, i.e. a random string independent from every $\pk_i \in \pp$ and known by every user;
    
    \item $n$: the expected number of players active in each step of the protocol;
    
    \item $\tH=\lfloor\frac{2n}{3}\rfloor+1$: a threshold used in the protocol, derived from the expected lower bound of the number of honest players in each step;
    
    \item $m$: the number of components of the list of arbitrary values upon which agreement has to be achieved, and which is common knowledge since the nodes know the events they must observe and describe on the ledger;
    
    \item $V = \prod_{c = 1}^m V_c$: the set the list to be agreed upon belongs to, each set $V_c$ contains all the possible values of the $c$-th component of said list;
    
    \item $\bot$: a special value that represents a \emph{non meaningful value} for any component, we require that $\bot \in V_c$ for every $1\le c \le m$;
    
    \item $\oo\in V$: the list built by player $i$ at the start of the protocol, for every $1\le c\le m$ we will say that $c$ is an \emph{unambiguous component} if $O_{i, c} = O_{j, c}$ for every couple of honest users $i, j$, otherwise, if there exist two honest users $i$ and $j$ such that $O_{i, c} \ne O_{j, c}$, we will say that $c$ is an \emph{ambiguous component};  
    
    \item $\Omega$: the amount of time spent by each player $i$ at the start of the protocol to build its private list $\oo$ (e.g. by observing some events and reporting some relevant information about them), the protocol will then try to reconcile all these lists  into a shared one;

    \item$\Lambda$: the upper bound to the time needed to propagate the messages in each of the first two steps of the protocol; 
    
    \item $\lambda$: the upper bound to the time needed to propagate the messages of the third and following steps of the protocol.
    The difference between $\lambda$ and $\Lambda$ depends on the size of the elements of $V$.
    We assume that $\Lambda=\mathcal{O}\left(\lambda\right)$;
    
    \item $s\in\Z^+$: the current step of the protocol;
    
    \item $\p s$: the active players that partake in step $s$ of the protocol;
    
    \item $\sigma_i^{\left(s\right)}=S\left(\sk_i,H\left(r\|s\right)\right)$: the credential of user $i$ for step $s$, used to check if $i\in\p s$;
    
    \item $p\in\left(0,1\right)$: for each time-slot $s$, each user in $\pp$ is chosen to be in $\p s$ with probability $p=\frac{n}{N}$;
    
    \item $\Mp s$ and $\hp s$: they are respectively the set of malicious and honest players in step $s$, note that $\Mp s\cup \hp s=\p s$ and $\Mp s\cap \hp s=\emptyset$;
    
    \item $\mm i s$: the message broadcast by player $i$ during step $s$;

    \item $\vv i s$: the list of information contained in the message $\mm i s$, we will see that $\vv i s \in V$ if $s \le 2$, $\vv i s \in \{0,1\}^m$ if $s \ge 3$;
    
    \item $\sig i s \left(x\right) = \left(x, S\left(\sk_i, s\|x\right)\right)$: the value $x$ broadcast by player $i$ during step $s$ certified by its signature, it is included in $\mm i s$;
    
    \item $C_i$: the certificate built by player $i$ which attests that the final list has network agreement, each user $i$ continues running the protocol until it can build a certificate $C_i$;
    
    \item $\alpha_i \in [0, \lambda]$: the time at which user $i$ starts the execution of each step of the protocol;
    
    \item $\beta_i^{\left(s\right)}$: the time at which user $i$ ends the execution of step $s$ of the protocol;
    
    \item $t^{\left(s\right)}$: the amount of time that players of step $s$ have to wait in order to harvest all the information required to compute the message to broadcast, if $i \in \p s$ then $t^{\left(s\right)} = \beta_i^{\left(s\right)} - \alpha_i$;
    
    
    \item $T_c$: the time at which the first honest user finalizes component $c$;
    
    \item $T$: the time at which the first honest user produces a certificate;
    
    \item $L$: a random variable representing the number of Bernoulli trials needed to see the output $1$, when each trial outputs $1$ with probability $\frac{h}{2}$;
    
    \item $\chi_{l,\frac{h}{2}}$: a random variable representing the number of steps required to end the probability game described in \cite{flamini2021multidimensional} with parameters $l$ and $\frac{h}{2}$. The probability game consists into flipping $l$ distinct but equal coins (which flip heads with probability $\frac{h}{2}$) until each of them flipped head at least once. Its probability distribution is computed in \cite{flamini2021multidimensional} but is also reported in \Cref{expected_steps}.
\end{itemize}

\subsection{Cob Protocol Description}
\labelx{protocol description}
We now describe in detail how Cob works.
The honest users will be the ones who follow the protocol described below, and, even if not elected as players on any step, they are supposed to stay online to support message propagation during the whole protocol execution.

The protocol is a variant of the MBA, where the first three steps are essentially the Multidimensional Graded Consensus presented in~\cite{flamini2021multidimensional}, then from the step $4$ onward it is a three-step loop that corresponds to the Multidimensional Binary Byzantine Algorithm, also presented in~\cite{flamini2021multidimensional}.

As in the protocol MBA, each user $i$ privately saves a list $\ff$ initialized to $\zeros$ that keeps track of the finalization of the components.
A component is finalized when the network is in agreement on it, and from that moment on the protocol will not change it anymore.
Once every component has been finalized, the protocol enables the creation of certificates that attest that the list is indeed shared by the network, and then terminates.
That is, the three-step loop is repeated until the ending condition is met, which corresponds to the creation of a certificate for the agreed-upon final list.

Every honest user $i$ in the system starts the protocol execution when its own private clock signs~0.
Note that, right from the start, each user $i$ can build its credentials $\sigma_i^{\left(s\right)}$ and check for which $s$ it will be $\phi\big(H\big(\sigma_i^{(s)}\big)\big) \le p$ and therefore $i\in\p s$.

We now describe Cob, followed by the Ending Condition to be performed in each step $s\ge 4$ to determine whether agreement has been achieved.
\pagebreak
\step{1}{first step of $m$-dimensional GC}
\begin{itemize}
    \item Each user $i$ computes its credential $\sigma_i^{\left(1\right)}$ and checks if $i\in\p 1$;
    
    \item if $i\not\in \p{1}$ then $i$ ends its step 1 right away;
    
    \item if $i\in \p{1}$, $i$ spends $t^{\left(1\right)}=\Omega$ time building its own private list $\oo \in V$, then:
    \begin{itemize}
        \item sets $\vv i 1 = \oo$;
        \item broadcasts the message:
        $$\mm i 1 =\left(1, \sigma_i^{\left(1\right)}, \sig i 1\left(\vv i 1\right)\right)\;.$$
    \end{itemize}
\end{itemize}

\vspace*{-10pt}
\step{2}{second step of $m$-dimensional GC}
\begin{itemize}
    \item Each user $i$ computes its credential $\sigma_i^{\left(2\right)}$, if $i\not\in \p{2}$ then $i$ ends its step 2;
    \item if $i\in \p{2}$, after waiting an amount of time $t^{\left(2\right)}=t^{\left(1\right)}+\Lambda+\lambda$, player $i$ does the following:
    \begin{itemize}
        \item sets $\vv i 2$, where $v_{i,c}^{\left(2\right)}=v_c\ne\bot$ if and only if $\#_i^{\left(1\right)}\left(v_c,c\right)\ge \tH $, and $v_{i,c}^{\left(2\right)}=\bot$ otherwise;
        \item broadcasts the message:
        $$\mm i 2 =\left(2,\sigma_i^{\left(2\right)},\sig i 2\left(\vv i 2\right)\right)\;.$$
    \end{itemize} 
\end{itemize}

\step{3}{output determination of $m$-dimensional GC and starting broadcast in $m$-dimensional BBA}
\begin{itemize}
    \item Each user $i$ collects and locally saves the messages received from the players of step 2;
    \item after waiting an amount of time $t^{\left(3\right)}=t^{\left(2\right)}+\lambda+\Lambda$, $i$ updates its private list $\oo$ and computes the list $\gc$ where, for each component $c$, $O_{i,c}$ and $g_{i,c}$ are computed as follows:
		\begin{itemize}
			\item if, for some $x \ne \bot$, $\#_i^{\left(2\right)}\left(x,c\right)\ge \tH$, then $\left(O_{i,c},g_{i,c}\right)=\left(x,2\right)$;
			\item else, if, for some $x \ne \bot$, $\#_i^{\left(2\right)}\left(x,c\right)\ge \frac{\tH}{2}$, then $\left(O_{i,c},g_{i,c}\right)=\left(x,1\right)$;
			\item otherwise, $\left(O_{i,c},g_{i,c}\right)=\left(\bot,0\right)$;
		\end{itemize}
	
    \item if $i\not\in \p 3$, then $i$ ends the execution of step 3;
    
    \item if $i\in \p{3}$, then player $i$ does the following:
    \begin{itemize}
        \item builds the list $\vv i 3 \in \{0, 1\}^m$ such that $v_{i,c}^{\left(3\right)}=0$ if $g_{i,c}=2$, $v_{i,c}^{\left(3\right)}=1$ otherwise;
	
    	\item computes the list $\TT i 3$ such that $\Theta_{i,c}^{\left(3\right)}=\bot$ if $v_{i,c}^{\left(3\right)}=1$, $\Theta_{i,c}^{\left(3\right)}=O_{i,c}$ when $v_{i,c}^{\left(3\right)}=0$;
    	
        \item broadcasts the message:
        $$\mm i 3=\left(3,\sigma_i^{\left(3\right)},\sig i 3\left(\vv i 3\right), \sig i 3\left(H\left(\TT i 3\right)\right)\right)\;.$$
    \end{itemize}
\end{itemize}

\step{$\mathbf{s}$ \hspace{5pt} $4\le s$, $s-1 \equiv 0\mod 3$}{Coin-Fixed-To-0 step and starting broadcast of Coin-Fixed-To-1 step in $m$-dimensional BBA}
\begin{itemize}
    \item Each user $i$ collects the messages received from the players active during step $s-1$;
    
    \item after waiting an amount of time  $t^{\left(s\right)}=t^{\left(s-1\right)}+2\lambda$, the user $i$ starts building the list $\vv i s$ performing the following operations:
    \begin{itemize}
        \item verifies the ENDING CONDITION;
	    
	    \item sets $v_{i,c}^{\left(s\right)}=v_{i,c}^{\left(s-1\right)}$ for all $1\le c \le m$ such that $f_{i,c}=1$;
	    
	    \item performs the FINALIZATION CHECK 0;
	    
	    \item performs the FINALIZATION CHECK 1; 
    \end{itemize}
    
    \item if $i\not\in \p s$, the user $i$ ends the execution of step $s$;
    
    \item if $i\in \p s$, the player $i$ does the following:
	\begin{itemize}    
        \item  completes the list $\vv i s$ depending on the lists $\vv j {s-1}$ included in the valid messages it has received, in particular, for each component $c$ such that $f_{i,c}=0$:
        \begin{itemize}
            \item if $\#_i^{\left(s-1\right)}\left(1,c\right)\ge \tH$, then $i$ sets $v_{i,c}^{\left(s\right)}=1$;
            \item else $i$ sets $v_{i,c}^{\left(s\right)}=0$;
        \end{itemize}
  
        \item computes the list $\TT i s$ such that $\Theta_{i,c}^{\left(s\right)}=\bot$ when $v_{i,c}^{\left(s\right)}=1$, $\Theta_{i,c}^{\left(s\right)}=O_{i,c}$ when $v_{i,c}^{\left(s\right)}=0$;
       
        \item broadcasts the message:
        $$\mm i s=\left(s,\sigma_i^{\left(s\right)},\sig i s\left(\vv i s\right), \sig i s\left(H\left(\TT i s\right)\right)\right)\;.$$
   \end{itemize}
\end{itemize}

\step{$\mathbf{s}$ \hspace{5pt} $5\le s$, $s-1 \equiv 1\mod 3$}{Coin-Fixed-To-1 step and starting broadcast of Coin-Genuinely-Flipped step in $m$-dimensional BBA}
\begin{itemize}
    \item Each user $i$ collects the messages received from the players active during step $s-1$;
    
    \item after waiting an amount of time  $t^{\left(s\right)}=t^{\left(s-1\right)}+2\lambda$, the user $i$ starts building the list $\vv i s$ performing the following operations:
    \begin{itemize}
        \item verifies the ENDING CONDITION;
	    
	    \item sets $v_{i,c}^{\left(s\right)}=v_{i,c}^{\left(s-1\right)}$ for all $1\le c \le m$ such that $f_{i,c}=1$;
	    
	    \item performs the FINALIZATION CHECK 0;
	    
	    \item performs the FINALIZATION CHECK 1; 
    \end{itemize}
    
    \item if $i\not\in \p s$, the user $i$ ends the execution of step $s$;
    
    \item if $i\in \p s$, the player $i$ does the following:
	\begin{itemize}    
        \item  completes the list $\vv i s$ depending on the lists $\vv j {s-1}$ included in the valid messages it has received, in particular, for each component $c$ such that $f_{i,c}=0$:
        \begin{itemize}
            \item if $\#_i^{\left(s-1\right)}\left(0,c\right)\ge \tH$, then $i$ sets $v_{i,c}^{\left(s\right)}=0$;
            \item else $i$ sets $v_{i,c}^{\left(s\right)}=1$;
        \end{itemize}
  
        \item computes the list $\TT i s$ such that $\Theta_{i,c}^{\left(s\right)}=\bot$ when $v_{i,c}^{\left(s\right)}=1$, $\Theta_{i,c}^{\left(s\right)}=O_{i,c}$ when $v_{i,c}^{\left(s\right)}=0$;
       
        \item broadcasts the message:
        $$\mm i s=\left(s,\sigma_i^{\left(s\right)},\sig i s\left(\vv i s\right), \sig i s\left(H\left(\TT i s\right)\right)\right)\;.$$
   \end{itemize}
\end{itemize}

\step{$\mathbf{s}$ \hspace{5pt} $6\le s$, $s-1 \equiv 2\left(\mod 3\right)$ }{Coin-Genuinely-Flipped step and starting broadcast of Coin-Fixed-To-0 step in $m$-dimensional BBA}
\begin{itemize}
    \item Each user $i$ collects the messages received from the players active during step $s-1$;
    
    \item after waiting an amount of time  $t^{\left(s\right)}=t^{\left(s-1\right)}+2\lambda$, the user $i$ starts building the list $\vv i s$ performing the following operations:
    \begin{itemize}
        \item verifies the ENDING CONDITION;
	    
	    \item sets $v_{i,c}^{\left(s\right)}=v_{i,c}^{\left(s-1\right)}$ for all $1\le c \le m$ such that $f_{i,c}=1$;
	    
	    \item performs the FINALIZATION CHECK 0;
	    
	    \item performs the FINALIZATION CHECK 1; 
    \end{itemize}
    
    \item if $i\not\in \p s$, the user $i$ ends the execution of step $s$;

    \item if $i\in \p s$, the player $i$ does the following:
	\begin{itemize}   
    
    \item  completes the list $\vv i s$ depending on the lists $\vv j {s-1}$ included in the valid messages it has received, in particular, for each component $c$ such that $f_{i,c}=0$:
        \begin{itemize}
            \item if $\#_i^{\left(s-1\right)}\left(0,c\right)\ge \tH$, then $i$ sets $v_{i,c}^{\left(s\right)}=0$;
            \item if $\#_i^{\left(s-1\right)}\left(1,c\right)\ge \tH$, then $i$ sets $v_{i,c}^{\left(s\right)}=1$;
            \item otherwise, letting $\p{s-1}_i \subseteq \p{s-1}$ be the set of players who sent $i$ a valid message in the previous step, then $i$ sets $v_{i,c}^{\left(s\right)}=k_c$, where $k=H\left(\min_{j \in \p{ s-1}_i}H\left(\mult{\sigma}_j^{\left(s-1\right)}\right)\right)$;
        \end{itemize}
         we will refer to the player whose hashed credential is minimal from $i$'s point of view as the \emph{coin flipper} selected by $i$ during step $s$;
    \item computes the list $\TT i s$ such that $\Theta_{i,c}^{\left(s\right)}=\bot$ when $v_{i,c}^{\left(s\right)}=1$, $\Theta_{i,c}^{\left(s\right)}=O_{i,c}$ when $v_{i,c}^{\left(s\right)}=0$;
       
    \item broadcasts the message:
    $$\mm i s=\left(s,\sigma_i^{\left(s\right)},\sig i s\left(\vv i s\right), \sig i s\left(H\left(\TT i s\right)\right)\right)\;.$$
    
    \end{itemize}
\end{itemize}

\checkstep{ENDING CONDITION}
If, while user $i$ waits for the end of the current step (step $s$), there exist a string $\theta \in \{0,1\}^d$ and a step $s'$ such that:
\begin{itemize}
    \item $4\le s'$ with $s'-1 \equiv 0 \mod 3$;
    
    \item user $i$ has received at least $\tH$ messages $\mm j {s'-1}$ containing the signature of $s'-1\|\theta = s'-1\|H\left(\TT{j}{s'-1}\right)$ and at least $\tH$ messages $\mm j {s'}$ containing the signature of $s'\|\theta = s'\|H\left(\TT{j}{s'}\right)$;
\end{itemize}
then $i$ can build its certificate $C_i$, diffuse it in the network and terminate its execution of the protocol.

For the collision resistance of $H$ we can assume that there is a list $\TTT\in V$ such that $\theta = H\left(\mathbf{\Theta}\right)$, i.e. $i$ has collected $2\tH$ signatures that refer to the same list $\TTT$.
\Cref{lemmino} states that in this case $i$ has received, for each $1 \le c \le m$, at least a message $\mm j 2$ with $v_{j,c}^{\left(2\right)}=\Theta_c$, hence it can reconstruct the list $\TTT$.
Let ${\hatp{s-1}} \subseteq \p{s-1}$ and ${\hatp s} \subseteq \p s$ be the the sets of players (each of cardinality at least $\tH$) that have sent messages with the signature of $H\left(\TTT\right)$ as stated above, then $i$ can build its certificate as:
{\small
\begin{equation*}
    C_i = \left(\TTT, s, \left\{\left(\sig{j}{s-1}\left(H\left(\TTT\right)\right), \sigma_j^{\left(s-1\right)}\right)\right\}_{j \in {\hatp{s-1}}}, \left\{\left(\sig{j}{s}\left(H\left(\TTT\right)\right), \sigma_j^{\left(s\right)}\right)\right\}_{j \in {\hatp{s}}}\right)\;.
\end{equation*}
}

\checkstep{FINALIZATION CHECK 0}       
Let $s'$ be a step such that $4\le s' \le s$  and $s'-1\equiv 0 \mod 3$ (that is, step $s'$ is a Coin-Fixed-To-0 step).

For each component $c\in\{1,\dots,m\}$ such that $f_{i,c}=0$, if, considering the lists $\vv{j}{s'-1}$ contained in the valid messages $\mm{j}{s'-1}$ received by $i$, we have that $\#_i^{\left(s'-1\right)}\left(0,c\right)\ge \tH$, then $i$ sets:
\begin{itemize}
    \item $v_{i,c}^{\left(s\right)}=0$;
    \item $f_{i,c}=1$.
\end{itemize}

\checkstep{FINALIZATION CHECK 1}       
Let $s'$ be a step such that $4\le s' \le s$  and $s'-1\equiv 1 \mod 3$ (that is, step $s'$ is a Coin-Fixed-To-1 step).

For each component $c\in\{1,\dots,m\}$ such that $f_{i,c}=0$, if, considering the lists $\vv{j}{s'-1}$ contained in the valid messages $\mm{j}{s'-1}$ received by $i$, we have that $\#_i^{\left(s'-1\right)}\left(1,c\right)\ge \tH$, then $i$ sets:
\begin{itemize}
    \item $v_{i,c}^{\left(s\right)}=1$;
    \item $f_{i,c}=1$.
\end{itemize}

\vspace*{10pt}
\noindent\rule{\textwidth}{1pt}
\vspace*{10pt}

User $i$ keeps following the protocol instruction until the ending conditions are satisfied and $i$ is able to build a certificate $C_i$.
A certificate $C_i$ contains the list $\TTT$ on which the network has reached agreement, alongside a set of digital signatures for $H\left(\TTT\right)$ together with a proof that those who signed were indeed players of a specific step.
In particular $C_i$ contains at least $\tH$ signatures from players of a Coin-Genuinely-Flipped step $s$ and $\tH$ signatures from players of the subsequent subsequent Coin-Fixed-To-0 step $s+1$.

\section{Security Analysis}
\labelx{security}
We now outline the security proof for the protocol Cob.

\noindent
In the main theorem, \Cref{mainT}, we determine:
\begin{enumerate}
    \item an upper bound to the time needed by the first honest node to produce a certificate for the agreed upon list $\TTT$;
    \item the time interval when every honest user gets to know $\TTT$.
\end{enumerate} 

In order to prove this, in \Cref{prelRes} we show some preliminary results characterising the gossiping communications under our assumptions, and  prove a lemma that justifies the construction of a certificate as described in \Cref{protocol description}.

In \Cref{componentwise} we prove some propositions and lemmas regarding the time needed by a network of nodes to reach consensus on the single list components.
In particular we distinguish two cases:
\begin{enumerate}
    \item if the nodes observed unambiguous events (i.e. the honest nodes agree on the same value at the beginning of the protocol) or very ambiguous events (which means that there is not a majority of the nodes who observed the same value), then they will reach $c$-agreement within step 5;
    \item otherwise, the number of steps required to reach $c$-agreement is upper bounded by $3L+5$, where $L$ is a Bernoulli-like random variable with parameter $\frac{h}{2}>\frac{1}{3}$.
\end{enumerate}

Finally, we use the results above to prove that the nodes of the network will be in possess of a certificate for the agreed upon list $\TTT$ within a number of steps upper-bounded by $5+3\chi_{\ell,\frac{h}{2}}$, and that only one list can be certified.

\begin{thm}\labelx{mainT}(Main Theorem)
Given an instance of Cob, described in \Cref{protocol description}, 
the following properties about each protocol execution hold with overwhelming probability:
\begin{enumerate}
    \item if $0 \le \ell \le m$ is the number of ambiguous components, then we have that $T \le \Omega + 2 \Lambda + \left(7 + 6 \chi_{\ell,\frac{h}{2}}\right) \lambda$, where $\chi_{\ell,\frac{h}{2}}$ is the random variable described in \cite{flamini2021multidimensional};
   
    \item all honest users agree on the same list $\TTT$ and know it in the interval ${[T,T+\lambda]}$.
\end{enumerate}
\end{thm}

\begin{remark}
When there are no ambiguous components, i.e. all honest nodes at the beginning of the protocol build the same list, then we have $\chi_{\ell,\frac{h}{2}} = 0$, so ${T \le \Omega + 2 \Lambda + 7 \lambda}$.
\end{remark}

\subsection{Preliminary Results}
\labelx{prelRes}
In order to prove our Main \Cref{mainT}, we first need to prove some preliminary lemmas and propositions which characterize Cob under our communication and network model.

\begin{lem} \labelx{prel}
For each step $s\ge1$ of a protocol run we have:
\begin{enumerate}
    \item \labelx{bb} if $i\in \p s$ is honest, then $\beta_i^{\left(s\right)}\in[t^{\left(s\right)},t^{\left(s\right)}+\lambda]$;
    
    \item \labelx{cc} if $i\in \p s$ is honest, then by time $\beta_i^{\left(s\right)}$ it has received all messages sent by all honest players $j\in \hp{s'}$ for all steps $s'<s$;

    \item \labelx{dd} for each step $\bar{s}> s$, fixing a component $1\le c\le m$, with overwhelming probability there do not exist two players $i,i' \in \p {\bar{s}}$ such that:
    \begin{itemize}
        \item $i$ has received at least than $\tH$ messages $\mm{j}{s}$ advertising $v_{j,c}^{\left(s\right)}=b$;
        \item $i'$ has received at least than $\tH$ messages $\mm{j}{s}$ advertising $v_{j,c}^{\left(s\right)}=b'$ with $b'\ne b$.
    \end{itemize}
    Note that for step 2 and 3 we have $b,b'\in V_c$, while for the next steps $b,b'\in \{0,1\}$.
\end{enumerate}
\end{lem}
\begin{proof}
Property \ref{bb} holds as a consequence of the network assumptions regarding the same speed clocks delay.
In fact we know that each user $i$ starts its protocol execution at a time $\alpha_i\in [0,\lambda]$ and waits for time $t^{\left(s\right)}$ before acting and then ending its step.
This means that $\beta_i^{\left(s\right)}=\alpha_i + t^{\left(s\right)} \in[t^{\left(s\right)},t^{\left(s\right)}+\lambda]$.\\

Property \ref{cc} holds by the definition of the protocol, noticing that $t^{\left(s\right)} \ge t^{\left(s'\right)}$ for all $s'<s$.
If $s=2,3$, then for all $s'<s$, $i \in \p s$, we have that
\begin{equation*}
    \beta_i^{\left(s\right)}\ge t^{\left(s\right)}=t^{\left(s-1\right)}+\lambda+\Lambda\ge t^{\left(s'\right)}+\lambda + \Lambda \ge \beta_j^{\left(s'\right)}+\Lambda,
\end{equation*}
since the honest players $j\in\hp{s'}$ send their messages of step $s'$ at time $\beta_j^{\left(s'\right)}$ and the messages reach all honest users in at most $\Lambda$ time, then player $i$ has received all the messages from honest players of the previous steps.\\
If $s\ge4$, then:
\begin{equation*}
    \beta_i^{\left(s\right)}\ge t^{\left(s\right)}=t^{\left(s-1\right)}+2\lambda \ge t^{\left(s'\right)}+\lambda+\lambda \ge \beta_j^{\left(s'\right)}+\lambda,
\end{equation*}
since each honest player $j\in\hp{s'}$ sends its message of step $s'$ at time $\beta_j^{\left(s'\right)}$, then it will reach all honest players by time $\beta_j^{\left(s'\right)}+\lambda \le \beta_i^{\left(s\right)}$.\\

Finally we prove Property \ref{dd}.
Let us assume for sake of contradiction that the two players $i,i'$ and the two values $b,b'$ of Property \ref{dd} do exist.
Note that each malicious player $j\in \Mp{s} $ may have signed both a list $\vv{j}{s}$ with $v_{j,c}^{\left(s\right)}=b$ and another list $\mathbf{v'}_j^{\left(s\right)}$ with ${v'}_{j,c}^{\left(s\right)}=b'$, but all honest players have signed exactly one list, hence their $c$-th component is unequivocal.

Therefore, at least $\tH-\Mp{s}$ of the messages received by $i$ advertising $b$ must come from a set $\mult{H}$ of honest players, and $\tH-\Mp{s}$ must come from another set $\mult{H}'$ of honest players with $\mult{H}$ and $\mult{H}'$ disjoint sets.
Note that the messages advertising different values in the $c$-th component must be distinct messages, this means that we are considering at least $2\tH$ distinct messages. 

Let $M$ be this set of at least $\tH$ messages collected by $i$ and $M'$ the analogous set of messages collected by $i'$, then:
\begin{equation*}
    2\tH\leq|M|+| M'| \le |\mult{H}| + | \Mp{s}| + |\mult{H}'| + |\Mp{s}| \le |\hp{s}| + 2|\Mp{s}|<2\tH,
\end{equation*}
where the last inequality holds with overwhelming probability thanks to the assumptions of \Cref{th_assumption}.
This is a contradiction, therefore such players $i$ and $i'$ do not exist.\flushright\qed
\end{proof}

Now we prove a lemma that justifies the construction of a certificate as described in \Cref{protocol description}.
In particular, the messages which constitute a valid certificate do not contain the list that the network is certifying, but just its digest.
Therefore a node must be able to determine which is the list associated to that digest.
In \Cref{lemmino} we prove that a node can find the candidate values for each components from the messages it has received in step 2.

\begin{lem}
    \labelx{lemmino}
    If a user $i$ builds a certificate $C_i$ for the list $\TTT$, then, for each $1\le c\le m$, $i$ has received at least one step 2 message from $j\in \hp 2$ with $v_{j,c}^{\left(2\right)}=\Theta_c$.
    \end{lem}
\begin{proof}
Note that from the assumptions of \Cref{th_assumption} on the number of malicious players we have that ${|\Mp s |< \frac{2\tH - |\hp s|}{2} < \frac{2\tH - \tH}{2} = \frac{\tH}{2}}$, so at least one of the signatures in the certificate $C_i$  must come from an honest player $k\in \hp s $.
Then $k$ must have received, during step 3, at least $\frac{\tH}{2}$ messages for $\Theta_c$ in $c$-th component.
Again, since $| \Mp s | < \frac{\tH}{2}$, at least one of them must come from an honest player $j\in\hp 2$ and, according to \Cref{cc} of \Cref{prel}, his message must have reached also $i$ within time $\beta_i^{\left(3\right)}$.

To conclude, note that $\beta_i^{\left(3\right)}$ is the ending time of step 3 for player $i$ and it is before any possible certificate production time.
\flushright\qed
\end{proof}

\subsection{Component-Wise Finalization}
\labelx{componentwise}
In this section we prove some properties about the finalization of a single component, distinguishing between the associated ambiguous and unambiguous events to be recorded.

We recall that the finalization checks are performed after every step $s \ge 4$ and refer to messages exchanged during step $s'\ge3$, and $s'\equiv 0 \mod 3$ (i.e. STEP 3 and all subsequent Coin-Genuinely-Flipped steps) for what concerns FINALIZATION CHECK 0 and step $s''\ge 4$, and $s''\equiv 1 \mod 3$ (i.e. all Coin-Fixed-To-0 steps)for what concerns FINALIZATION CHECK 1.

\subsubsection{Unambiguous Components}
In the following proposition we will show how the network behaves if the event associated to a specific component is unambiguous. In particular, we will explain, following the protocol steps, why every honest player will finalize that component within the end of STEP 5.
\begin{prop}($c$-Agreement on Unambiguous Components)
\labelx{unambiguous}
Let $c$ be an unambiguous component, then the following happens with overwhelming probability:
\begin{itemize}
    \item all honest users have their $c$-th component finalized by step $5$ (and in particular there is $c$-agreement on the lists $\TT i s$ for all $s\ge 5$);
    \item $T_c\le t^{\left(5\right)}+\lambda$.
\end{itemize}
\end{prop}

\begin{proof}
Note that every honest player $i\in \hp s $ starts its step $s$ at time $\alpha_i \in [0, \lambda]$.
Now we analyse the protocol step by step.
\begin{description}
    \item[\textbf{STEP 1}]
    Since component $c$ is unambiguous, then for a certain value $x\in V_c$ each honest player $i\in \hp{1}$ will build a list $\vv i 1$ with $v_{i,c}^{\left(1\right)}=x$.
    Then $i$ will propagate its message $\mm i 1$ at time $\beta_i^{\left(1\right)}= \alpha_i + \Omega$.
    
    \item[\textbf{STEP 2}]
    When an honest player $i\in \hp{2}$ stops waiting at time ${\beta_i^{\left(2\right)}=\alpha_i+t^{\left(2\right)}}$, $i$ has received all step 1 messages sent by the other honest players.
    
    By our assumptions we have, with overwhelming probability, ${| \hp{1} | >\tH}$, hence more than $\tH$ step 1 messages $\mm j 1$ that $i$ has received contain a list $\vv j 1$ with $v_{j,c}^{\left(1\right)}=x$.
    Then, whether $x=\bot$ or $x\ne\bot$, player $i$ builds a list $\vv i 2$ with $v_{i,c}^{\left(2\right)}=x$ and broadcasts the message $\mm{i}{2}$ containing the digital signature of this list.
    
    \item[\textbf{STEP 3}]
    When an honest player $i\in \hp{3}$ stops waiting at time ${\beta_i^{\left(3\right)}=\alpha_i+t^{\left(3\right)}}$, $i$ has received all step 2 messages from all the honest players. 

    Since the lists in their messages $\mm{j}{2}$ have $v_{j,c}^{\left(2\right)}=x$, and with overwhelming probability $| \hp{2} | >\tH$, then player $i$ will set $\left(O_{i, c}, g_{i, c}\right)=\left(x,2\right)$ if $x\ne\bot$, $\left(O_{i, c}, g_{i, c}\right)=\left(x,0\right)$ if $x=\bot$.
    So, $i$ will build the list $\vv i 3$ with $v_{i,c}^{\left(3\right)}=0$ if $x\ne\bot$ or $v_{i,c}^{\left(3\right)}=1$ if $x=\bot$, and broadcast its message $\mm{i}{3}$.
    
    \item[\textbf{STEP 4}]
    When an honest player $i\in \hp{4}$ stops waiting at time ${\beta_i^{\left(4\right)}=\alpha_i+t^{\left(4\right)}}$, $i$ has received all step 3 messages from all the honest players. We now consider separately two cases:
    \begin{itemize}
        \item $x\ne\bot$, in this case player $i$ enters the FINALIZATION CHECK 0 ($4-1\equiv 0 \mod 3$) and since the number of honest players is $| \hp{3} | \ge \tH$ with overwhelming probability, player $i$ sets $v_{i,c}^{\left(4\right)}=0$ and $f_{i,c}=1$.
        This means that all honest players have finalized the $c$-th component of the list and they will get $\Theta_{i,c}^{\left(s\right)}=x$ for all $s\ge4$.
    
        \item $x=\bot$, in this case player $i$ will neither enter the ENDING CONDITION nor any FINALIZATION CHECK.
        It will build a list $\vv i 4$ such that $v_{i,c}^{\left(4\right)}=1$ since with overwhelming probability $| \hp{3} | >\tH$.
        Player $i$ will broadcast its message $\mm{i}{4}$ containing the digital signature of $\vv i 4$.
    \end{itemize}
    Thus $c$-agreement has been reached if $x\ne\bot$, otherwise we will see that it will be reached in the next step.
    
    \item[\textbf{STEP 5}]
    When an honest player $i\in \hp{5}$ stops waiting at time ${\beta_i^{\left(5\right)}=\alpha_i+t^{\left(5\right)}}$, $i$ has received all step 4 messages from all the honest players.
    Again, we consider two cases:
    \begin{itemize}
        \item $x\ne\bot$, in this case $c$-agreement on lists $\TT{i}{s}$ has already been reached, and the $c$-th component has already been finalized by the honest players.
        
        \item $x=\bot$, in this case, $i$ has received with overwhelming probability at least $\tH$ messages from all the other honest players $j\in \hp{4}$ containing the digital signature of a list $\vv{j}{4}$ with $v_{j,c}^{\left(4\right)}=1$.
        Then $i$ enters the FINALIZATION CHECK 1 and sets $v_{i,c}^{\left(5\right)}=1$ and $f_{i,c}=1$.
        This means that all honest players have finalized the $c$-th component of the list and they will get $\Theta_{i,c}^{\left(s\right)}=\bot$ for all $s\ge5$.
    \end{itemize}
    Since the malicious players are less than $\tH$, they will not be able to produce the number of messages required to mislead the honest players.
    We have seen that by the end of step $5$ all honest users have finalized the $c$-th component (and they are in agreement with each other), so we have that $T_c\le t^{\left(5\right)}+\lambda$. \flushright\qed
\end{description}
\end{proof}

\subsubsection{Ambiguous Components}
Let us now tackle the more difficult case of ambiguous components.
In \Cref{ambiguous1} we deal with the simpler sub-case, when  no honest player sets $g_{i,c} = 2$ during step 3, which means that there is a wide disagreement among the network about that specific component.
As we will see, this case will resolve with the achievement of agreement on the symbol $\bot$.

Then, we complete the analysis by considering the case when some honest node sets $g_{i,c} = 2$.
In \Cref{GC_case2} we prove that in this case each honest user $j$ has saved the same value $O_{j,c}=x\in V_c$ at the beginning of step 3.
They do not know yet that they already are in agreement, so each of them tries to figure out whether to preserve that component of the final list or to discard it by setting it to $\bot$. 
This decision will be made by exchanging the bit lists from step 3 onward.
In fact, as it is stated in \Cref{fromBinaryToGeneral}, once $c$-agreement is reached on the bit list either on 0 or 1, it is also reached on the list $\mathbf{\Theta}_j \in V$ respectively on $x\in V_c$ or $\bot$.

Then, it becomes essential to prove that $c$-agreement is achievable on each component with probability 1, and also to upper-bound the time required to achieve it.
To do that, we prove in \Cref{lem_coin_gen_flip} that the network will reach $c$-agreement on the bit list with probability greater than $\frac{1}{3}$ after every Coin-Genuinely-Flipped.
Therefore, in \Cref{lem_lungo} we prove that $c$-agreement is eventually reached with probability 1 and that all the honest nodes will finalize the component $c$ with the same value finalized by the first node in the network who can do it (even if the first user who can finalize the component is malicious). 

These Lemmas are then used to prove \Cref{ambiguous2}, where we present an upper bound to the number of steps and time required to finalize a single component.
Finally, the results of the previous lemmas and propositions are used to prove \Cref{mainT}.

\begin{prop}[$c$-Agreement on Very Ambiguous Components]
\labelx{ambiguous1}
Let $c$ be an ambiguous component and assume that all honest step 3 players set $g_{i,c}<2$, then with overwhelming probability we have that:
\begin{itemize}
    \item all honest users have their $c$-th component finalized (in particular will be in $c$-agreement on the list $\TT i s$) in step 5, setting $\Theta_{i,c}^{\left(s\right)}=\bot$;
    \item $T_c\le t^{\left(5\right)}+ \lambda$.
\end{itemize}
\end{prop}
\begin{proof}
    By definition of the protocol Cob, for each $i\in \hp{3}$, $i$ sets $v_{i,c}^{\left(3\right)}=1$, since $g_{i,c}<2$.
    This means that the honest step 3 players start in agreement on the component $c$ of the list $\vv i 3$.
    They may not be in agreement on the list component $O_{i,c}$, but this does not matter. 
    In fact, during step 4 no honest player is able to finalize component $c$ by collecting more than $\tH$ messages with $v_{k,c}^{\left(3\right)}=0$, since $| \Mp{3} | < \tH$.
    In the same way even if the honest players have received more than $\tH$ valid messages with $v_{k,c}^{\left(3\right)}=1$ they will not finalize the component $c$ because $4-1\not\equiv 1 \mod 3$.
    Anyway, each honest step 4 player $i$ has received all honest messages, hence more than $\tH$ advertising $v_{k,c}^{\left(3\right)}=1$.
    This means that $i$ will create a message $\mm{i}{4}$ with $v_{i,c}^{\left(4\right)}=1$.
    
    During step 5, which is a Coin-Fixed-To-1 step, each honest user $i$ will receive all messages by other honest players before $\beta_i^{\left(5\right)}$, hence with overwhelming probability $i$ will receive $\tH$ messages for $v_{k,c}^{\left(4\right)}=1$.
    Thus $i$ will enter the FINALIZATION CHECK 1 and will finalize the $c$-th component by setting $v_{i,c}^{\left(5\right)}=1$ and $f_{i,c}=1$.
    
    This means that the honest users reach $c$-agreement on $\Theta_{i,c}^{\left(s\right)}=\bot$ by the end of step $5$ (which happens in the interval $[t^{\left(5\right)}, t^{\left(5\right)} + \lambda]$), and it will hold for all $s\ge 5$. 
    Note that in this case, a strong disagreement among the players results in the component to be set to $\bot$ as it happens in \Cref{unambiguous} when every honest players starts with $\bot$.
    \vspace*{-10pt}\flushright\qed
\end{proof}

\begin{prop}[$c$-Agreement on Ambiguous Components]
\labelx{ambiguous2}
Let $c$ be an ambiguous component and assume that there exists a player $i\in\hp 3$ which sets $g_{i,c}=2$, then:
\begin{itemize}
    \item all honest users have their $c$-th component finalized (in particular will be in $c$-agreement on the list $\TT i s$) within step $3L+5$;
    \item $T_c\le t^{\left(3L+5\right)}+\lambda$.
\end{itemize}
Where $L$ is a random variable representing the number of Bernoulli trials needed to see the output~$1$, when each trial outputs $1$ with probability $\frac{h}{2}$.
\end{prop}

In order to prove \Cref{ambiguous2} we need the results stated in the following four lemmas.

\begin{lem}
\labelx{GC_case2}
Under the assumption of \Cref{ambiguous2}, we show that the following properties hold:
    \begin{enumerate}
        \item \labelx{p1} $g_{j,c}\ge1$ for all $j\in \hp{3}$;
        \item \labelx{p2}there is a value $x\in V_c$ such that $O_{j,c}=x$ for all $j\in \hp{3}$.
    \end{enumerate}
\end{lem}

\begin{proof}

Since player $i\in \hp{3}$ is honest and sets $g_{i,c}^{\left(3\right)}=2$, then:
\begin{enumerate}
    \item $i$ sets $O_{i,c}=x$ since it has received more than $\tH$ messages $\mm{k}{2}$ advertising $v_{k,c}^{\left(2\right)}=x$.
    By Property 3 of \Cref{prel} we know that no honest player $j\in \hp{3}$ has received $\tH$ messages $\mm{k}{2}$ for $v_{k,c}^{\left(2\right)}=x'\ne x$, hence if $g_{j,c}^{\left(3\right)}=2$ it must be $O_{j,c}=x$.
    
    As showed in the proof of \Cref{lemmino}, we have that with owerwhelming probability
    $|\Mp s |<\frac{\tH}{2}$,
    so we can state that more than $\frac{\tH}{2}$ honest players must have signed for $x$.
    Therefore, if $g_{j,c}^{\left(3\right)} < 2$, then $g_{j,c}^{\left(3\right)}=1$, and Property \ref{p1} holds.

    \item We now show that, even if $j$ sets $g_{j,c}=1$, it will set $O_{j,c}=x$.
    In fact, there can not exist a value $x'\ne\bot$ and $x'\ne x$ such that $j$ has received also more than $\frac{\tH}{2}$ step 2 messages $\mm k 2$ with $v_{k,c}^{\left(2\right)}=x'$.
    For the sake of contradiction, we suppose that these messages exist; many of them may come from malicious players in $\Mp{2}$, but at least one of them must come from an honest player $p\in \hp{2}$.
    This means that $p$ has received $\tH$ step 1 messages $\mm k 1$ with $v_{k,c}^{\left(1\right)}=x'$.
    Since we have seen that some other honest step~2 players have signed a step~2 message advertising $v_{k,c}^{\left(2\right)}=x$, this implies that they have seen $\tH$ step~1 messages with $v_{k,c}^{\left(1\right)}=x$, which, by \Cref{prel}, is a contradiction. This means that Property~2 holds. 
\end{enumerate}\flushright\qed
\end{proof}

\begin{lem}
\labelx{fromBinaryToGeneral}
Under the assumptions of \Cref{ambiguous2} we have that
$c$-agreement on the list $\TTT$ is reached when $c$-agreement is reached on the bit list $\vv{}{}$.
\end{lem}

\begin{proof}
Since the honest player $i\in\hp{3}$ sets $g_{i,c}=2$, it will set $v_{i,c}=0$, therefore it is possible that $c$-agreement is reached on 0.
By the analysis of Property 2 of \Cref{GC_case2} the honest users may not have an agreement on their $v_{i,c}^{\left(3\right)}$ at the end of step 3 but they will have an agreement on $O_{i,c}^{\left(3\right)}=x$.
This means that, by the definition of Cob, when $c$-agreement is reached on the bit list (either on 0 or 1), then it will also be reached on $x$ in $\TTT$ (respectively on $x$ or $\bot$). \flushright\qed
\end{proof}

\begin{remark}
This property is true also in the general case, but we have explicitly proved it only under the assumptions of \Cref{ambiguous2}.
\end{remark}

\begin{lem}
\labelx{lem_coin_gen_flip}Being $c$ a component on which agreement among the honest user does not hold, at every Coin-Genuinely-Flipped step $c$-agreement is reached with probability at least $\frac{h}{2}$.
\end{lem}
\begin{proof}
Assuming that $c$-agreement is not reached at the beginning of a Coin-Genuinely-Flipped step $s$ where $s\ge 6$, $s-1 \equiv 2 \mod 3$, let an honest player $i\in \hp s $ be in the condition that it must flip the coin during such step.

This means that the player $i$ has not received more than $\tH$ messages for a bit $b\in\{0,1\}$ in component $c$, so $i$ selects its own coin flipper $\ell'$, and $i$ will set $v_{i,c}^{\left(s\right)}=b_i=H\left(H\left(\sigma_{\ell'}^{\left(s-1\right)}\right)\right)_c$, where $H\left(K\right)_c$ is the $c$-th bit of $H\left(K\right)$.
    
Note that, by \Cref{prel}, if an honest player has seen more than $\tH$ messages for the same bit $b$ in component $c$, then no honest player has seen more than $\tH$ messages for $1-b$ in the same component.
This means that $i$ will be in $c$-agreement with the honest players who did not flip the coin only if $b_i=b$, and this happens with probability $\frac{1}{2}$ with the Random Oracle assumption for the hash function $H$.

Therefore, all honest players in $\hp s $ will be in agreement with probability $\frac{1}{2}$.
Actually, this is true if the coin flipper is an honest player, in fact in this case $b_i$ can be assumed to be randomly chosen and globally shared among the honest players flipping the coin. 
If the player with minimal hashed credential is a malicious player, then we cannot say much about the probability distribution of the output of the bit extraction, since some players may not have seen its message.
    
However, with our assumptions on the common reference string $r$, we can state that, with probability $h$, the coin flipper will be honest, and in this case all honest players will be in agreement with probability $\frac{1}{2}$.

Combining these two independent probabilities we get that with probability at least $\frac{h}{2}$ the honest players reach $c$-agreement every time they enter a Coin-Genuinely-Flipped step.
Note that they will finalize this component within the following 2 steps: in $s+1$ if $b_i=0$, in $s+2$ if $b_i=1$. \flushright\qed
\end{proof}

\begin{lem}
\labelx{lem_lungo}
Under the assumptions of \Cref{ambiguous2}, we have that: 
\begin{enumerate}
    \item being $E$ the event ``there exists a step $\hat{s}\ge 4$ such that, for the first time, some user $\hat{\imath} \in \pp$ (either malicious or honest) should finalize its $c$-th component of list $\vv{\hat{\imath}}{\hat s}$'', $E$ happens with probability 1;
    
    \item $T_c\le t^{\left(\hat{s}+3\right)} + \lambda$ and $c$-agreement is reached in step $\hat{s}$ on the same value finalized by $\hat{\imath}$ (however, the $c$-th component might be finalized 3 steps later).
\end{enumerate}
\end{lem}
\begin{proof}
\begin{enumerate}
    \item 
    As proven in \Cref{lem_coin_gen_flip}, if $c$-agreement is not reached by the honest users, at every Coin-Genuinely-Flipped step they will reach it with probability at least $\frac{h}{2}>\frac{1}{3}$.
    
    Therefore, once every 3 steps the $c$-th component will be finalized with probability greater than $\frac{1}{3}$, therefore the probability that the event \emph{E} happens converges to 1.
    
    \item Step $\hat{s}$ is the first step in which a user $\hat{\imath}$ can finalize the $c$-th component. By the construction of the protocol, this happens in two possible ways:
    \begin{description}
        \item[$E_a$:] $\hat{\imath}$ is able to collect or generate (and then propagate) at least $\tH$ valid messages $\mm{k}{s'-1}$ with $v_{k,c}^{\left(s'-1\right)}=0$, $4 \le s' \le \hat{s}$, and $s' - 1 \equiv 0 \mod 3$;
        
        \item[$E_b$:] $\hat{\imath}$ is able to collect or generate (and then propagate) at least $\tH$ valid messages $\mm{k}{s'-1}$ with $v_{k,c}^{\left(s'-1\right)}=1$, $4 \le s' \le \hat{s}$, and $s' - 1 \equiv 1 \mod 3$;
    \end{description}
    Because the messages produced during step $s'-1$ by honest players are received by every user before they are done waiting in step $s'$, and because the adversary receives everything no later than the honest users, without loss of generality we can assume that $s' = \hat{s}$, and that the user $\hat{\imath}$ is malicious.
    
    For any step $s\ge 4$,
    every honest player $i \in \hp s $ who has waited time $t^{\left(s\right)}$ has received all honest step $s-1$ messages (thanks to \Cref{prel}), and all honest players in $\hp s $ have set $O_{i,c}=x$ (according to \Cref{GC_case2}).
    
    We now consider step $\hat{s}$ and examine 4 exhaustive ways in which event \emph{E} may happen.
    \begin{description}
        \item[Case 2.1.a:] \emph{event $E_a$ happens and there is an honest user $i' \in \pp$ who should also finalize the $c$-th component}.
        
        In this case, we have $\hat{s} - 1 \equiv 0 \mod 3$, hence Step $\hat{s}$ is a Coin-Fixed-To-0 step.
        By assumption, $i'$ has received at least $\tH$ valid step $\hat{s} - 1$ messages $\mm{k}{\hat{s} - 1}$ with $v_{k,c}^{\left(\hat{s} - 1\right)}=0$.
        Thus $i'$ finalizes its component $c$ setting $v_{i',c}^{\left(\hat{s}\right)}=0$ and $f_{i',c}=1$.
        
        Now we show that any other honest user $i$
        has either finalized its $c$-th component, setting $v_{i,c}^{\left(\hat{s}\right)}=0$ and $f_{i,c}=1$, or
        has set $v_{i,c}^{\left(\hat{s}\right)}=0$ without finalizing such component.
        
        Because step $\hat{s}$
        is the first time any player $i$ should finalize component $c$ of the list $\vv{i}{s}$, there does not
        exist a Coin-Fixed-To-1 step $s' < \hat{s}$ (hence $s'-1 \equiv 1 \mod 3$) such that $\tH$ players have signed $v_{i,c}^{\left(s'-1\right)}=1$.
        Accordingly, no online user in $\pp$ finalizes the list component $c$ in step $s'$ setting $v_{i,c}^{\left(s'\right)}=1$.
        Moreover, if an honest user has waited for a time $t^{\left(\hat{s}\right)}$, then it must have received all step $\hat{s}-1$ messages from  honest players, and (considering the messages received by $i'$) at least $\tH-| \Mp{\hat{s}-1} | \ge 1$ must have $v_{k,c}^{\left(\hat{s}-1\right)}=0$.
        According to Property 4 of \Cref{prel}, an honest player $i$ cannot collect $\tH$ messages with $v_{k,c}^{\left(\hat{s}-1\right)}=1$, therefore it sets $v_{i,c}^{\left(\hat{s}\right)}=0$.
        
        For step $\hat{s}+1$, since user $i'$ has helped propagating the messages that have let it finalize the $c$-th component on or before time $\alpha_{i'}+ t^{\left(\hat{s}\right)}$, then on or before time $\beta_i^{\left(\hat{s}+1\right)}$ each honest user $i$ has received at least $\tH$ valid $ \hat{s}-1$ messages for the bit 0.
        In fact, even if some of the $\tH$ messages received by $i'$ were not broadcast in time by a malicious user, within time $\beta_i^{\left(\hat{s}+1\right)}$ they have reached $i$, for all $i \in \pp$.
        This is true because user $i'$ received the messages within time $\beta_{i'}^{\left(\hat{s}\right)}$ and helped propagating them, hence within time $\beta_{i'}^{\left(\hat{s}\right)}+\lambda$ they have reached all honest players, and for all honest players $i\in \hp{\hat{s}+1}$ we have:
        \begin{equation*}
            \beta_{i}^{\left(\hat{s}+1\right)} \ge t^{\left(\hat{s}\right)} + 2\lambda\ge \alpha_{i'} + t^{\left(\hat{s}\right)}+\lambda = \beta_{i'}^{\left(\hat{s}\right)} + \lambda.
        \end{equation*}
        Furthermore, honest players will not end step $\hat{s}+1$ before receiving those step $\hat{s}-1$ messages, because there do not exist other $\tH$ valid step $s'-1$ messages for 1 in the component $c$ with $s'- 1 \equiv 1 \mod 3$ and $5 \le s' < \hat{s} + 1$, by the definition of Step $\hat{s}$ in assumption $E_a$ (step $\hat{s}$ is the first step in which a user should finalize component $c$).
        In particular, step $\hat{s} + 1$ itself is a Coin-Fixed-To-1 step, but no honest player has propagated during step $\hat{s}$ a message for 1 (as we have shown they have reached $c$-agreement on 0), and $| \Mp{\hat{s}}| < \tH$.
        Thus all honest users finalize their $c$-th component, setting $v_{i,c}^{\left(\hat{s}+1\right)}=0$ and $f_{i,c}=1$.
        
        So, we have proven that, for all $s\ge \hat{s} +1$, the honest users set $v_{i,c}^{\left(s\right)}=0$ and $f_{i,c}=1$ (i.e. they finalize the component $c$ within the end of step $\hat{s}+1$).
        We have already seen that $\exists x \in V_c$ such that $O_{i,c} = x$ for every honest user $i$, so they will set $\Theta_{i,c}^{\left(s\right)} = x$ and therefore $c$-agreement reached on $\TT{i}{s}$ for all $s\ge \hat{s} +1$.
        
        \item[Case 2.1.b:] \emph{event $E_b$ happens and there is an honest user $i' \in \pp$ who should also finalize $c$-th component}.
        
        In this case we have $\hat{s}-1\equiv 1 \mod 3$, then step $\hat{s}$ is a Coin-Fixed-To-1 step.
        The analysis is similar to \emph{Case 2.1.a} and we will omit many details.
        
        As in the previous case, $i'$ must have received $\tH$ valid step $\hat{s}-1$ messages with $v_{k,c}^{\left(\hat{s}-1\right)}=1$.
        Again, by the definition of step $\hat{s}$, there does not exist a step $s'$, with $4\le s'\le \hat{s}$ and $s'-1\equiv 0 \mod 3$, where at least $\tH$ players have signed a message with $v_{k,c}^{\left(s'-1\right)}=0$.
        Thus, $i'$ finalizes the $c$-th component and sets $v_{i',c}^{\left(\hat{s}\right)}=1$ and $f_{i',c}=1$.
        Moreover, any other honest user $i\in \pp$ has either finalized its $c$-th component if it has received $\tH$ messages with $v_{k,c}^{\left(\hat{s}-1\right)}=1$ or has set $v_{i,c}^{\left(\hat{s}\right)}=1$ and broadcast its $\mm{i}{\hat{s}}$ message.
        Since $i'$ has helped propagating the step $\hat{s}-1$ messages it has received by time $\alpha_{i'}+t^{\left(\hat{s}\right)}$, all honest users finalize the $c$-th component during step $\hat{s}+1$.
        
        Again, they will reach $c$-agreement over $\bot$ on $\TT{i}{s}$ for all $s\ge \hat{s}$, they will set $v_{i,c}^{\left(\hat{s}+1\right)}=1$ and $f_{i,c}=1$, finalizing the component $c$ within the end of step $\hat{s}+1$.
        
        \item[Case 2.2.a:] \emph{event $E_a$ happens and there does not exist an honest user ${i' \in \pp}$ who should also finalize $c$-th component}.
        
        In this case, note that $\hat{\imath}$ could have received or generated $\tH$ step $\hat{s}- 1$ messages with $v_{k,c}^{\left(s-1\right)}=0$.
        However, the malicious users may not help propagating those messages, so we cannot conclude that the honest
        users will receive them after time $\lambda$.
        In fact, $| \Mp{\hat{s}-1} |$ of those messages may be from malicious players, who did not propagate their messages at all and only sent them to the other malicious players cooperating with them.
        
        Therefore, the honest users will wait for time $t^{\left(\hat{s}\right)}$ without finalizing component $c$. However, by Property 4 of \Cref{prel}, they will not see more than $\tH$ of the messages received with $v_{k,c}^{\left(\hat{s}-1\right)}=1$, again because with overwhelming probability $| \hp{\hat{s}-1} | + 2| \Mp{\hat{s}-1} | < 2\tH$.
        Since step $\hat{s}$ is a Coin-Fixed-To-0 step, every honest player $i\in \hp{\hat{s}}$ thus sets $v_{i,c}^{\left(\hat{s}\right)}=0$ and propagates its message at time $\alpha_i + t^{\left(\hat{s}\right)}$.
        
        During step $\hat{s} + 1$, which is a Coin-Fixed-To-1 step, two things may happen:
        \begin{enumerate}
            \item [1] an honest user receives the $\tH$ messages received by $\hat{\imath}$ (who decided to propagate them and let it finalize component $c$): in this case the situation is similar to \emph{Case 2.1.a}, and every honest user $i$ will finalize its $c$-th component within time $\alpha_i+ t^{\left(\hat{s}+1\right)}+ \lambda$;
            
            \item [2] the honest users will receive at least $| \hp{\hat{s}} |$ ($>\tH$ with overwhelming probability) messages with $v_{k,c}^{\left(\hat{s}\right)}=0$ from the honest players.
            Then they propagate their messages with $v_{i,c}^{\left(\hat{s}+1\right)}=0$ but do not finalize since step $\hat{s}+1$ is not a Coin-Fixed-To-0 step.
            
            In this case, in step $\hat{s}+2$ which is a Coin-Genuinely-Flipped step, two things may happen:
        \begin{enumerate}
            \item [2.1.] if $\hat{\imath}$ broadcasts the step $\hat{s}-1$ messages that let it finalize its $c$-th component, then the honest users will finalize their $c$-th component as well setting $v_{i,c}^{\left(\hat{s}+2\right)}=0$ and $f_{i,c}=1$, hence reaching $c$-agreement over $\Theta_{i,c}^{\left(s\right)}=x$ for all $s\ge \hat{s}+2$ within time $\alpha_i+t^{\left(\hat{s}+2\right)}$;
            
            \item [2.2.] otherwise all honest users have received all step $\hat{s}+1$ messages from the honest players with $v_{i,c}^{\left(\hat{s}+1\right)}=0$. Again they are more than $\tH$, so the honest users set $v_{i,c}^{\left(\hat{s}+2\right)}=0$ without flipping the coin.
            Again, they do not finalize component $c$ since $\hat{s}+2$ is not a Coin-Fixed-To-0 step so they just broadcast their step $\hat{s}+2$ messages.
            
            Finally, step $\hat{s}+3$ is a Coin-Fixed-To-0 step, so everyone will receive at least $\tH$ messages with $v_i^{\left(\hat{s}+2\right)}=0$ where $i\in \hp{\hat{s}+2}$.
            Then all honest users $k$ at time $\alpha_k+t^{\left(\hat{s}+3\right)}$ can finalize component $c$ setting $v_{k,c}^{\left(\hat{s}+3\right)}=0$ and $f_{k,c}=1$ and hence reach $c$-agreement over $\TT{k}{s}=x$ for all $s\ge \hat{s}+3$.
        \end{enumerate}
        \end{enumerate}
        
        Depending on how $\hat{\imath}$ and in general the malicious users behave, some users may finalize the component $c$ within the end of step $s$ (with ${s\in\{\hat{s}, \hat{s}+1, \hat{s}+2\}}$) using step $\hat{s}-1$ messages, or within the end of step $\hat{s}+3$ with step $\hat{s}+2$ messages.
        It does not matter since $c$-agreement is reached anyway over $\TT{i}{s}$ for all $s\ge \hat{s}$, the component $c$ is finalized setting $f_{i,c}=1$ within step $\hat{s}+3$, and  $v_{i,c}^{\left(s\right)}=0$ for every step $s$ such that $s\ge \hat{s}+3$.
        
        \item[Case 2.2.b:] \emph{event E.b happens and there does not exist an honest user ${i' \in \pp}$ who should also finalize $c$-th component}.
        
        The analysis in this case is similar to \emph{Case 2.1.b} and \emph{Case 2.2.a}, thus many details have been omitted.
        
        We know that $\hat{\imath}$ has collected or generated at least $\tH$ step $\hat{s}-1$ messages with $v_{k,c}^{\left(\hat{s}-1\right)}=1$ and $\hat{s}-1 \equiv 1 \mod 3$ (hence $\hat{s}$ is a Coin-Fixed-To-1 step) and that no honest player could have seen more than $\tH$ messages for 0.
        Thus each honest player $i\in \hp{\hat{s}}$ sets $v_{i,c}^{\left(\hat{s}\right)}=1$ and propagates its message $\mm{i}{\hat{s}}$ at time $\alpha_i+t^{\left(\hat{s}\right)}$.
        Similar to \emph{Case 2.2.a}, within 3 steps user $i$ will finalize their $c$-th component.
        
        Then $c$-agreement is reached over $\TT{i}{s}$ for all $s\ge \hat{s}$, and the component $c$ is finalized setting $f_{i,c}=1$ within step $\hat{s}+3$ and  $v_{i,c}^{\left(s\right)}=0$ for all steps $s$, $s\ge \hat{s}+3$.
    \end{description}
    Combining the four sub-cases, we obtain:
    \begin{itemize}
        \item  $T_c\le t^{\left(\hat{s}\right)}+\lambda$ in \emph{Case 2.1.a} and \emph{Case 2.1.b};
        \item $T_c\le t^{\left(\hat{s}+3\right)}+\lambda$ in \emph{Case 2.2.a} and \emph{Case 2.2.b};
    \end{itemize}
    but we also have that they all are in agreement at the end of step $\hat{s}$, and that $c$-agreement is reached over $\TT{i}{s}$ for all $s\ge \hat{s}$.
\end{enumerate}
\flushright\qed
\end{proof}

Now we can prove \Cref{ambiguous2}.
\begin{proof}[Proof of \Cref{ambiguous2}]
    Given the results of \Cref{lem_lungo}, it remains to upper-bound $\hat{s}$ and thus $T_c$.
    We do that by considering how many times the Coin-Genuinely-Flipped steps are executed by at least one honest player.
    
    If no honest player flips the coin in a Coin-Genuinely-Flipped step $s$, it means that they all have received more than $\tH$ messages with ${v_{k,c}^{\left(s-1\right)}=b\in\{0,1\}}$ and $c$-agreement has been reached, letting them finalize the component in at most 2 more protocol steps.
    Moreover if they reach $c$-agreement over $0$, this means that they agree on the same value $O_{k,c}$ to insert as $c$-th component of the list $\TT{k}{s}$.
    Once $c$-agreement is reached in step $s$, the honest players will finalize the $c$-th component either in step $s+1$ or step $s+2$ depending on whether $b=0$ or $b=1$.
    
    By \Cref{lem_coin_gen_flip}, at every Coin-Genuinely-Flipped step $c$-agreement is reached with probability at least $\frac{h}{2}$, so we can compare this step to a Bernoulli trial that outputs 1 if $c$-agreement is reached.
    This means that, before step $\hat{s}$ (the first step in which a user can finalize the $c$-th component), the distribution of the number of times the Coin-Genuinely-Flipped steps are executed to finalize a component $c$ can be upper-bounded by to the random variable $L$, which we recall represents the number of Bernoulli trials needed to see a 1 when each trial gives 1 with probability $\frac{h}{2}>\frac{1}{3}$.
    Letting $s'$ be the last Coin-Genuinely-Flipped step before the finalization of the $c$-th component, then we have, by the protocol construction, $s'=3+3L$.
    
    Assuming that the adversary knows the outcome of $L$ in advance, when should the adversary make step $\hat{s}$ happen to maximize the delay of the finalization time $T_c$ of the $c$-th component by an honest user?
    
    If $\hat{s}>s'$ (hence $ \hat{s}=s'+1$ or $ \hat{s}=s'+2 $) then this means that we are in \emph{Case 2.1.a} or \emph{Case 2.1.b} of \Cref{lem_lungo} since at the end of step $s'$ the honest players are already in agreement, so when a malicious player could finalize, also the honest users can, hence
    \begin{equation*}
        T_c\le t^{\left(\hat{s}\right)}+\lambda \le t^{\left(s'+2\right)}+\lambda.
    \end{equation*}
    
    If $\hat{s}<s'-3$, that is $\hat{s}$ is before the second to last Coin-Genuinely-Flipped step, then by the analysis of \emph{Case 2.2.a} or \emph{Case 2.2.b} we get
    \begin{equation*}
        T_c\le t^{\left(\hat{s}+3\right)}+\lambda \le t^{\left(s'\right)}+\lambda,
    \end{equation*}
    that is, the Adversary is making the agreement on component $c$ happen faster.
    
    If $\hat{s}=s'-1$ or $\hat{s}=s'-2$, then $\hat{s}$ is the Coin-Fixed-To-0 or Coin-Fixed-To-1 step before $s'$.
    By the analysis of the 4 sub-cases we know that the honest players never flip the coin and finalize the $c$ component within the next two steps.
    Therefore, the following holds:
    \begin{equation*}
        T_c\le t^{\left(\hat{s}+3\right)}+\lambda \le t^{\left(s'+2\right)}+\lambda.
    \end{equation*}
    
    To summarize, no matter what $\hat{s}$ is, we have:
    \begin{equation*}
        T_c\le t^{\left(s'+2\right)}+\lambda = t^{\left(3L+5\right)}+\lambda,
    \end{equation*}
    which upper-bounds the time needed to reach agreement on the $c$-th list component. \flushright\qed
\end{proof}

Now we will prove \Cref{mainT}, using the results of the previous lemmas and propositions.
We will prove that all the honest users will agree on the same $\TTT$, that no malicious user can build a valid certificate for a different $\hat{\TTT}$, and that the honest users will be able to produce a certificate for $\TTT$ within time $t^{\left(5 + 3 \chi_{\ell,\frac{h}{2}}\right)}+\lambda$.
We also prove that only one list can be certified, therefore the nodes will reach agreement on a list and it is not possible for the malicious users to produce a valid certificate for another list. This guarantees the consistency property of Cob.

\begin{proof}[Proof of \Cref{mainT}]
It is sufficient to note that, since $t^{\left(s+1\right)} = t^{\left(s\right)} + 2\lambda$ for all $s \ge 3$, we have that:
\begin{align}
    t^{\left(5 + 3 \chi_{\ell,\frac{h}{2}}\right)} &= t^{\left(1\right)} + \sum_{s = 2}^{5 + 3 \chi_{\ell,\frac{h}{2}}} \left(t^{\left(s\right)} - t^{\left(s-1\right)}\right)\\
    &= t^{\left(1\right)} + \left(t^{\left(2\right)} - t^{\left(1\right)}\right) + \left(t^{\left(3\right)} - t^{\left(2\right)}\right) + \sum_{s = 4}^{5 + 3 \chi_{\ell,\frac{h}{2}}} \left(t^{\left(s\right)} - t^{\left(s-1\right)}\right) \nonumber\\
    &= \Omega + \Lambda + \lambda + \Lambda + \lambda + \sum_{s = 4}^{5 + 3 \chi_{\ell,\frac{h}{2}}} 2\lambda \nonumber\\
    &= \Omega + 2\Lambda + 2\lambda + \left(2 + 3 \chi_{\ell,\frac{h}{2}}\right) 2\lambda \nonumber\\
    &= \Omega + 2\Lambda + \left(6 + 6 \chi_{\ell,\frac{h}{2}}\right) \lambda.
\end{align}

Let  $m-\ell$ be the number of unambiguous components, we have shown in \Cref{unambiguous} that these components will reach agreement in at most 5 steps.
The same will happen for some of the $\ell$ ambiguous components according to \Cref{ambiguous1}, while the others, as shown in \Cref{ambiguous2}, will reach agreement within a number of steps whose distribution is upper-bounded by the random variable $3L+5$.

Cob runs until a certificate for a list $\TTT\in V$ is created, this happens \emph{no later} than the moment in which every component is finalized by the honest players, which happens once $c$-agreement is reached on each of the $\ell$ ambiguous components.

Since we have shown in \Cref{lem_coin_gen_flip} that with probability at least $\frac{h}{2}$ the honest players will reach $c$-agreement on a single component, and once agreement is reached it is maintained for the whole protocol run, then agreement will be reached in at most $\chi_{\ell,\frac{h}{2}}$ Coin-Genuinely-Flipped steps (accordingly to the analysis in \Cref{ambiguous2}, malicious users might speed the consensus process up!) where $\chi_{\ell,\frac{h}{2}}$ is the same random variable described in \cite{flamini2021multidimensional}.

Every honest user will be able to obtain a certificate for a block at the end of the two steps following the $\chi_{\ell,\frac{h}{2}}$-th Coin-Genuinely-Flipped step according to the definition and analysis of Cob shown above. Therefore it holds that $T \le t^{\left(5 + 3 \chi_{\ell,\frac{h}{2}}\right)} + \lambda$.\\

Now we show that if a certificate is created for the first time in step $s$ for a list $\TTT\in V$, then any certificate created will be for the same list $\TTT$.

Let step $s$ be the first step in which a user $k$ is able to collect a certificate $C_k$ for $\TTT$.
We recall that, by the assumptions in \Cref{th_assumption}, given two distinct players $i, j \in \p s$ of the same step, it is negligible the probability that $i$ collects $\tH$ messages for a list $\TT{i}{s-1}$ and $j$ collects $\tH$ messages for a distinct list $\TT{j}{s-1}$.

As before, we can assume that the user $k$ is malicious and the certificate is made of step $s-2$ and step $s-1$ messages, where step $s$ is a Coin-Fixed-To-1 step.

We distinguish two cases:
\begin{itemize}
    \item \emph{There is an honest user $k'$ who also can collect a certificate $C_{k'}$ for $\TTT$ in step $s$}.
    
    In this case $k'$ has propagated the messages which let it certify the list $\TTT$, hence all honest users will be in possess of a certificate (possibly a different one) for the same list $\TTT$, so the honest users will agree on the same list $\TTT$.
    Also, the honest users will end the protocol execution, so there is no chance that another certificate is produced in the following steps since with overwhelming probability $\Mp s <\tH \; \forall s$.
    
    \item \emph{There is no honest user $k'$ who also can collect a certificate $C_{k'}$ for $\TTT$ in step $s$}.
    
    In this case the honest users will keep executing the protocol until they can create a certificate for a block $\hat{\TTT}$ or until they receive from $k$ the messages that allowed $k$ to create the certificate in step $s$.
    
    We will show that if they do not receive the certificate from $k$, then they will build a certificate for $\hat{\TTT}=\TTT$.
    This guarantees that honest users will agree on the same list, since there are no two valid certificates around the network for two distinct lists.

    If the user $k$ has built a certificate for $\TTT$ in step $s$, it means that $k$ has received $\tH$ messages from step $s-2$ and step $s-1$ for $\TTT$.
    In particular, $k$ collected $\tH$ step $s-2$ messages for $0$ in every component $c$ such that $\Theta_c\ne \bot$.
    We will call $I_0=\{c: \Theta_c\ne \bot, 1 \le c \le m\}$.
    By the assumptions in \Cref{th_assumption}, no honest user has received at least $\tH$ messages from step $s-2$ that sponsors 1 in a component $c\in I_0$.
    This implies that all the honest players have sent in step $s-1$ a message with 0 in each component $c\in I_0$.
    This brings all the honest users in $c$-agreement on such components and it will keep holding in the following steps.
    
    We also know that the user $k$ has received $\tH$ messages from step $s-1$ for $\TTT$.
    This means that $k$ has received at least $\tH$ messages for 1 in every component $c$ such that $\Theta_c = \bot$.
    We will call $I_1=\{c: \Theta_c= \bot, 1 \le c \le m\}$.
    Again, by the assumptions in \Cref{th_assumption}, no honest user has received at least $\tH$ messages from step $s-1$ that sponsor 0 in a component $c\in I_1$. 
    Therefore, all the honest players send, in step $s$, a message with 1 in each component $c\in I_1$.
    This brings all the honest users in $c$-agreement on such components and it will keep holding in the following steps.
    
    Now we note that $I_0\cup I_1=\{1,\dots,m\}$, hence all the honest users are in agreement on all list components.
    Therefore, in step $s+3$, which is again a Coin-Fixed-To-1 step, they will be able to build a certificate for the list of relevant information $\TTT$ using their messages of step $s+1$ and step $s+2$.\flushright\qed
\end{itemize}
\end{proof}

\section{Performance Analysis}
\labelx{performance}
Given \Cref{general_problem} and the context of application of Cob, described in \Cref{stateofart} (i.e. a consensus protocol for incomplete networks with millions of nodes), we present a comparison which highlights the advantages of using the leaderless and parallel protocol Cob instead of executing $\ell$ instances of Algorand to achieve the same result.

For the evaluation we will consider the most relevant use case of Cob, namely the example in \Cref{intro} of a blockchain implementing sharding where the network must agree on the blocks created by each shard.

In our performance analysis we will compare the total amount of data that is broadcast by the nodes to let the network reach consensus with the two approaches. 
We quantify the amount of data broadcast by the network to reach consensus on a list of $\ell$ ambiguous events (which are the most problematic case, hence the worst case scenario) when the ratio of honest users is 80\% (i.e. $h=0.8$) and the expected number of players in each step is $n=4000$, this choice of parameters which reflects the one used in \cite{chen2019algorand}. 

First, we will determine the expected number of steps in which the elected players broadcast their messages before producing a certificate both for Cob and for Algorand.
Then, we will approximate the weight of the messages broadcast in both protocols, and finally we will compare the total amount of data broadcast as the parameter $\ell$, the number of ambiguous components, changes.

\subsection{Expected Number of Steps}
\labelx{expected_steps}
Starting with Cob, let us make explicit the probability distribution of the random variable $\chi_{\ell,\frac{h}{2}}$ used to upper-bound the number of steps needed to reach consensus.
In the analysis of protocol MBA  \cite{flamini2021multidimensional}, the probability distribution of $\chi_{\ell,\frac{h}{2}}$ is shown to be:
\begin{align*} 
\probability\left(\chi_{\ell,\frac{h}{2}}=w\right)=\left(1-\left(1-\frac{h}{2}\right)^{w}\right)^\ell - \left(1-\left(1-\frac{h}{2}\right)^{w-1}\right)^\ell,
\end{align*}
from which it is possible to compute the expected number of loops of Coin-Fixed-To-0, Coin-Fixed-To-1 and Coin-Genuinely-Flipped steps the network must execute if the adversary manages to delay as much as possible the consensus achievement.
\Cref{mainT} states that the number of steps needed to produce a certificate for the agreed upon list is $5+3\chi_{\ell,\frac{h}{2}}$, therefore the last step in which messages are broadcast is step $4+3\chi_{\ell,\frac{h}{2}}$.
This means that the expected number of protocol steps with message broadcasting that are needed to produce a certificate can be computed as:
\begin{align*}
    \expected\left[\text{Cob steps}_{\ell,\frac{h}{2}}\right]=4+3 \expected\left[\chi_{\ell,\frac{h}{2}}\right]=4+3 \sum_{w=1}^{\infty}w \probability\left(\chi_{\ell,\frac{h}{2}}=w\right).
\end{align*}

With Algorand, the expected number of loops required to bring the nodes to agreement is $\frac{2}{h}$.
Before entering the loop that performs the binary consensus (Binary Byzantine Agreement \cite{micali2016byzantine}), the nodes execute 4 extra steps (Graded Consensus \cite{feldman1997optimal}).
Finally, in the worst case, the nodes will be able to produce a certificate 2 steps after the last Coin-Genuinely-Flipped step.
This means that we must consider one more step with message broadcasting after the last Coin-Genuinely-Flipped step.
Therefore, the expected number of steps with message broadcasting required by Algorand is:
\begin{align*}
    \expected\left[\text{Alg steps}_{\frac{h}{2}}\right]=4+3\frac{2}{h}+1=5+\frac{6}{h}\;.
\end{align*}

This is the expected number of steps to reach consensus on a single event.
Note that the network must execute an instance of Algorand for each of the $\ell$ events they observe.

\subsection{Weight of the Messages}
We begin with Algorand: in the first step Algorand selects a small number of nodes (e.g. 35 potential leaders) who must broadcast their proposal, namely the hash of a block created by a shard plus a verifiable credential.
The following steps are expected to be performed by 4000 nodes who broadcast messages of 200 bytes as specified in \cite{chen2019algorand}.
Since in this use case the weight of the step 1 messages is close to the weight of the other steps, but the number of messages broadcast is much less, we can neglect it while determining the whole amount of broadcast data.
The messages created from step 4 onwards weigh around 200 bytes (see \cite{chen2019algorand}) and contain the signature of a single bit, the signature of a digest and a verifiable credential.
Finally, the messages created in step 2 and step 3, contain the signature of a digest (the digest of the block created by a shard) together with the verifiable credential of the selected player, therefore we approximate the weight of credential and signature to around 100 bytes, and add the weight of the digest, i.e. 32 bytes, for a total of 132~B.\\

With Cob, in the first two steps the nodes of the network broadcast a message containing the signature of the list of digests of the blocks created by the shards, together with the verifiable credential of the node who created the message.
The weight of such messages can be approximated, with the same reasoning as before, to $32\ell+100$ bytes.
In the following steps, the nodes produce messages containing the signature of a list of $\ell$ bits, the signature of a digest and a verifiable credential.
These messages weight around $\frac{\ell}{8}+ 200$ bytes, because they contain the same data as a message of a step $s$, with $s\ge 4$, of Algorand, once we substitute the single bit of Algorand with the $\ell$ bit list of Cob.

In \Cref{tab:table weight steps,tab:table weight phases} we summarize the weight of the messages for each step of both Algorand and Cob applied to the sharding use case. In \Cref{tab:table weight steps} the weights are associated to the single steps of each protocol, while in \Cref{tab:table weight phases} the weight of messages are associated to the corresponding protocol phase.

\vspace{5 pt}
\begin{table}
\centering
  \renewcommand{\arraystretch}{1.5}
\begin{tabularx}{0.8\textwidth} { 
  | >{\centering\arraybackslash}X 
  | >{\centering\arraybackslash}X 
  | >{\centering\arraybackslash}X | }
 \hline
  \textbf{step} &  \textbf{Cob} &  \textbf{Algorand} \\
 \hline
  step 1  &  $32\ell+100$ bytes  &  200 bytes ($\star$) \\
\hline
 step 2 &  $32\ell+100$ bytes &  132 bytes \\
 \hline
  step 3  &  $\frac{\ell}{8}+ 200$ bytes  &  132 bytes  \\
\hline
 following steps &  $\frac{\ell}{8}+ 200$ bytes &  200 bytes \\
 \hline
\end{tabularx}
\caption{\labelx{tab:table weight steps}
Weight of messages of each step in a single run of Cob and Algorand.
The symbol~($\star$) recalls that the first step of  Algorand will not be considered in our comparison.
The parameter $\ell$ is the number of events the network must observe.
The dimensions reported are derived  from the analysis of \cite{chen2019algorand}.}
\vspace{5 pt}

\begin{tabularx}{0.8\textwidth} { 
  | >{\centering\arraybackslash}X
  | >{\centering\arraybackslash}X 
  | >{\centering\arraybackslash}X | }
 \hline
  \textbf{phase} &  \textbf{Cob} &  \textbf{Algorand} \\
 \hline
  Leader selection  &  no leader &  200 bytes ($\star$) \\
\hline
 graded consensus &  $32\ell+100$ bytes &  132 bytes \\
 \hline
  binary agreement &  $\frac{\ell}{8}+ 200$ bytes  &  200 bytes  \\
\hline
 
\end{tabularx}
\caption{\labelx{tab:table weight phases}Weight of messages of each step in the corresponding phase of the Cob and Algorand protocols.
See also \Cref{tab:table weight steps}.
}
\end{table}

\subsection{Comparison}
We recall that the choice of using a different instance of Algorand for each list component comes from the considerations of \Cref{parallel} and \Cref{leaderless}.
Recall that Algorand is a leader-based consensus protocol, so it is preferable to have one leader (and one protocol instance) for each single event, in order to let consensus achievement on each single component to be independent from the others.
In fact, if we used Algorand on the whole list (only a single instance of Algorand), then disagreement on a single component would cause the whole list to be discarded.
Therefore, we proceed with the comparison between the execution of Cob and the multiple instances of Algorand.

\subsubsection{Algorand}
We can put the information of \Cref{tab:table weight steps} together and compute the amount of data broadcast in the network in a single Algorand run.
As we mentioned earlier, we will not consider the messages broadcast in the first step because their contribution is practically negligible.

We also recall that, since Algorand is a leader-based consensus protocol, a malicious leader might deliberately produce a proposal which finds all (or the majority) of the nodes in the network at odds.
In this case the consensus process drops to the symbol $\bot$ and a certificate is built with step 5 messages.
This means that even if the event the network observes is ambiguous, if a leader acts maliciously, agreement will be reached fast.
Therefore, since it is not possible to predict how a malicious leader will act, in our comparison we will consider the two border cases: the case when the leader acts honestly, and the case in which every malicious leader will broadcast a controversial message making the consensus drop to $\bot$ in 5 steps.
\begin{description}
    \item [\textbf{Case 1.}] If all leaders (i.e. the players of step 1) act honestly, the amount of data broadcast in the network in a single run is expected to be:
\begin{align*}
     \expected[\mathrm{weight}(\mathrm{Alg}_{1,h,n,\text{honest}})] & =
     n\cdot\left( 2 \cdot 132+\left(\expected\left[\text{Alg steps}_{\frac{h}{2}}\right]-3\right)200\right)\\
     & =n\cdot\left( 264+\left(2+\frac{6}{h}\right)\cdot200\right)\;.
\end{align*}

Multiplying this value by $\ell$ we obtain the amount of data broadcast in the network by the nodes to reach consensus on $\ell$ ambiguous events:
\begin{align*}
   \expected[\mathrm{weight}(\mathrm{Alg}_{\ell,h,n,\text{honest}})] = \ell\cdot n\cdot\left( 264+\left(2+\frac{6}{h}\right)\cdot200\right)\;.
\end{align*}

\item [\textbf{Case 2.}] If all malicious leaders cause the drop of their event data, the total weight slightly decreases.
Recall that, in Algorand, a leader is honest with probability at least $p_h=h^2(1+h-h^2)$ (see \cite{chen2019algorand}).
In this case every instance of the protocol whose leader is malicious will end after the broadcast of step 5 and the consensus drops to $\bot$.
This means that:
\begin{align*}
     &\hspace*{-.3cm}\expected[\mathrm{weight}(\mathrm{Alg}_{1,h,n,\text{drop}})] \\
     &\ge n\cdot\left( 132\cdot 2 + 200\cdot 2 + 200\cdot\left(\expected\left[\text{Alg steps}_{\frac{h}{2}}\right]-5\right)\cdot h^2\left(1+h-h^2\right)\right)\\
     &=n\cdot\left( 264 +200\cdot\left(2+6 h\left(1+h-h^2\right)\right)\right)\;.
\end{align*}

Again, multiplying by $\ell$ we obtain the total amount of data broadcast in the network:
\begin{align*}
     \expected[\mathrm{weight}(\mathrm{Alg}_{\ell,h,n,\text{drop}})]\ge \ell\cdot n\cdot\left( 264 +200\cdot\left(2+6 h\left(1+h-h^2\right)\right)\right)\;.
\end{align*}
\end{description}

\subsubsection{Cob }
With Cob, the amount of data broadcast in a single protocol run (which covers all the $\ell$ components) can be computed as:
\begin{align*}
    \expected[\mathrm{weight}(\mathrm{Cob}_{\ell,h,n})]=n\left(2(100+32\ell) + \left(\expected[\text{Cob steps}_{\ell,\frac{h}{2}}]-2\right)\left(\frac{\ell}{8}+200\right)\right).
\end{align*}

To give the idea of how Cob can outperform Algorand, we compute the amount of data broadcast by a network where the percentage of honest users is 80\% (i.e. $h=0.8$), and the expected number of players that must broadcast a message, in each step is $n=4000$ (recall that we are not considering the first step of Algorand).
\Cref{fig:graph_log,fig:graph_zoom} show the expected total amount of data broadcast for different values of the parameter $\ell$, the number of ambiguous components.

\begin{figure}
    \centering
    \includegraphics[scale=0.5]{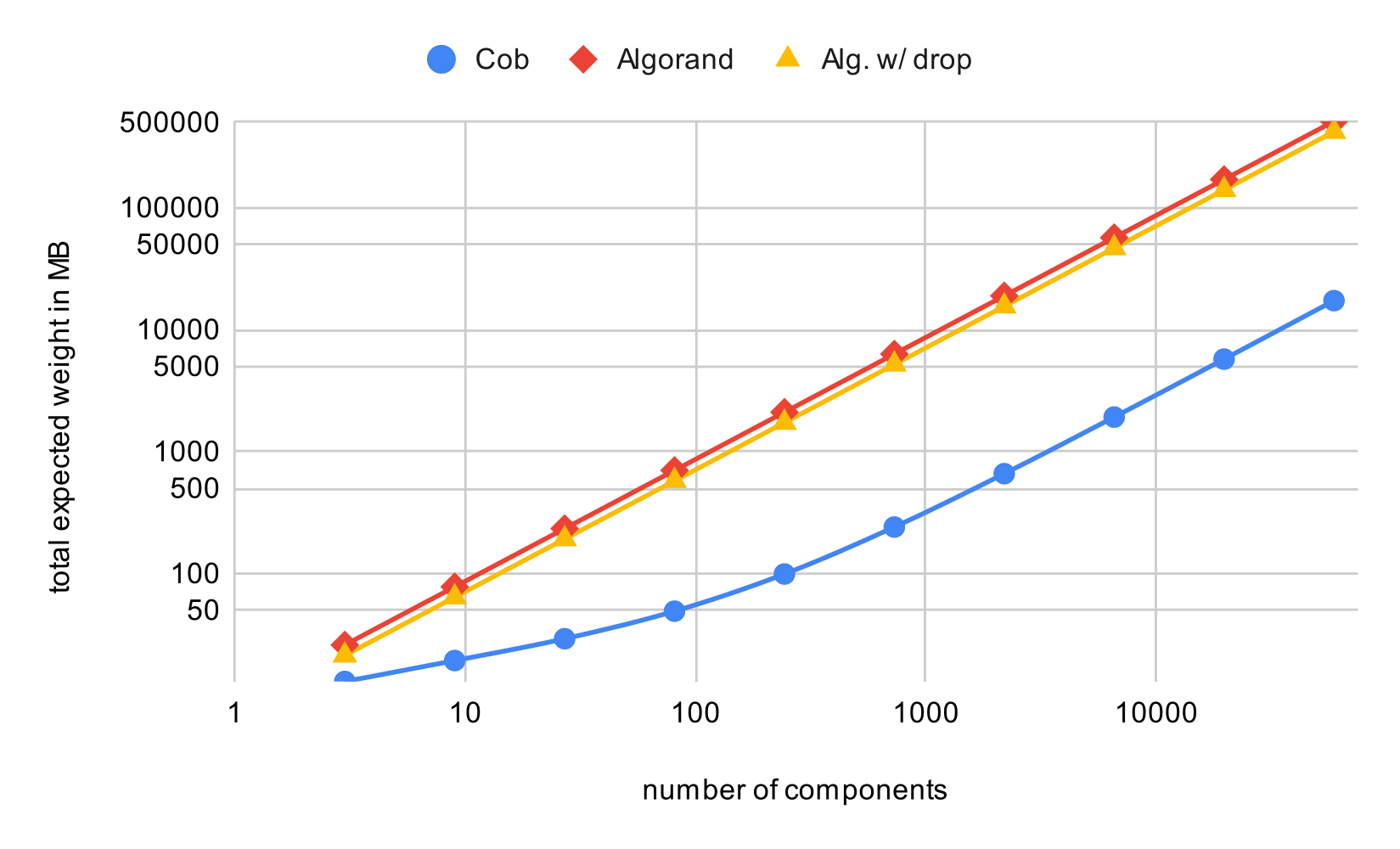}
    \caption{Amount of data broadcast in the network (in MB) using Algorand or Cob with parameters $h=0.8$ and $n=4000$ in terms of the number of components $\ell$, logarithmic scale in both axes.}
    \labelx{fig:graph_log}
    
    \includegraphics[scale=0.5]{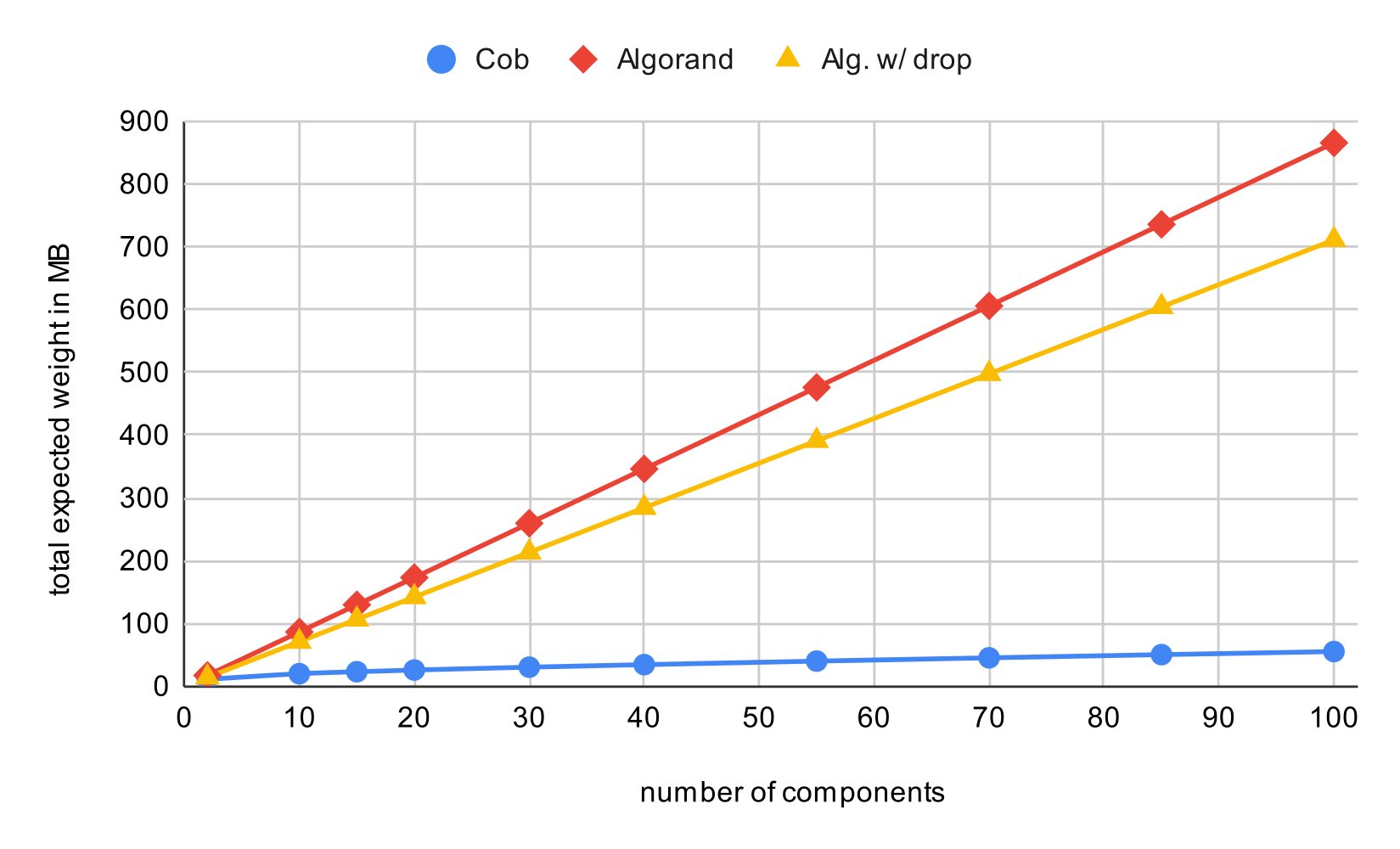}
    \caption{Amount of data broadcast in the network (in MB) using Algorand or Cob with parameters $h=0.8$ and $n=4000$ in terms of the number of components $\ell$, linear scale in both axes.}
    \labelx{fig:graph_zoom}
\end{figure}

We recall that, for the analysis of Algorand, we considered the two border cases, namely all the malicious leaders drop their components to $\bot$ and no malicious leader makes its component to  drop to $\bot$.
Therefore, in general, the number of MB broadcast in the network in any Algorand protocol execution will reasonably be between the two corresponding lines.

\section{Conclusions}
\labelx{conclusions}
We presented Cob, an extension of the MBA protocol \cite{flamini2021multidimensional} which allows the nodes of a wide gossiping network to reach consensus on a list of arbitrary values, working in parallel on each component.

This generalization widens the applications of the original protocol thanks to the sortition mechanism that limits the number of messages to be broadcast and processed at each step when there are many players, and the relaxed network assumptions which model real-case scenarios more closely.
Notice that the protocol retains the leaderless approach of the MBA protocol, a democratic feature that is valued in permissionless distributed settings and thwarts attacks from malicious leaders.
Moreover, it also preserves the parallel approach that enhances efficiency with respect to multiple executions of protocols designed to work in the same environment, such as Algorand.

As we explained in \Cref{intro}, we believe that one of the most relevant use cases of Cob is as consensus layer to allow the reconciliation of transactions in blockchain platforms implementing sharding.
In this context, we proposed a comparison between Cob and the execution of multiple instances of the well-known protocol Algorand, supposing that a network of nodes must reach consensus on which blocks have been legitimately created by different shards.
As shown in \Cref{fig:graph_log,fig:graph_zoom}, Cob remarkably reduces the amount of data broadcast in the network with respect to multiple executions of Algorand, and this would reasonably speed up the consensus process.

\subsection{Future Works}
Cob guarantees to reach consensus if the assumptions are met, and its leaderless and parallel approach maximizes the number of list components that are finalized on a meaningful value (i.e. $\ne \bot$).
However its execution is probabilistic, and although it halts with probability 1, the number of steps necessary to halt have only an upper bound in the form of a Bernoulli-like distribution.

An interesting research direction could focus on extending the protocol by introducing some \emph{termination steps}, in order to have a fixed upper bound on its execution, which would benefit many concrete applications.
Specifically, such an extension would see the protocol running normally up to a pre-determined number of steps, then, if the execution has not halted yet, the protocol starts a sequence of termination steps that guarantee to reach a consensus in a fixed number of steps.
In this phase it is quite tricky to try to preserve as much meaningful agreement as possible: the trivial solution is to collapse the agreement on $\bot$ if consensus is not reached in time, but avoiding to do so has to account for a wide array of attacks with which malicious players could try to disrupt agreement.

\renewcommand{\bibname}{References} 
\bibliographystyle{plain}
\bibliography{Bibliography}
\clearpage
\end{document}